\documentclass[a4paper,fleqn,usenatbib]{mnras}
\usepackage{graphicx}
\usepackage{epsfig}
\usepackage{amsmath}

\usepackage[T1]{fontenc}
\usepackage{ae,aecompl}
\usepackage{multicol}
\usepackage{booktabs}
\usepackage{amssymb}	
\usepackage{float}

\newcommand	\gtsim	{\lower.5ex\hbox{$\buildrel > \over \sim$}}
\newcommand     \ltsim  {\lower.5ex\hbox{$\buildrel < \over \sim$}}
\newcommand	\simgt	{\lower.5ex\hbox{$\buildrel > \over \sim$}}
\newcommand     \simlt  {\lower.5ex\hbox{$\buildrel < \over \sim$}}
\title[Formation of EL CVn-type Binaries]
{The Formation of EL CVn-type Binaries}

\author[Chen, Maxted, Li \& Han]
{Xuefei Chen$^{1,2,3}$\thanks{E-mail: cxf@ynao.ac.cn}, 
P.F.L. Maxted$^4$, Jiao Li$^{1,2,5}$, 
Zhanwen Han$^{1,2,3}$ \\
$^{1}$ Yunnan Observatories, Chinese Academy of Sciences, Kunming, 650011, China\\
$^{2}$ Key Laboratory for the Structure and Evolution of Celestial Objects, Chinese Academy of Sciences, Kunming, 650011, China \\
$^{3}$ Center for Astronomical Mega-Science, Chinese Academy of Sciences, 20A Datun Road, Chaoyang District, Beijing, 100012, China\\
$^{4}$ Astrophysics Group, Keele University, Keele, Staffordshire ST5 5BG, UK\\
$^{5}$ University of the Chinese Academy of Sciences, Yuquan Road 19, Shijingshan Block, 100049, Beijing, China
 }

\date{Accepted XXX. Received YYY; in original form ZZZ}

\pubyear{2016}

\begin{document}
\label{firstpage}
\pagerange{\pageref{firstpage}--\pageref{lastpage}}
\maketitle

\begin{abstract}
EL CVn-type binaries are eclipsing binaries that contain an A- or F-type dwarf star 
and a very low mass ($\sim 0.2M_\odot$) helium white dwarf precursor (proto-He WD).
A number of such objects have been discovered in the WASP and Kepler 
photometric surveys. Here we have studied the formation of EL CVn-type binaries 
and give the properties
and the space density of this population of stars in the Galaxy.
We show that EL CVn binaries cannot be
produced by common envelope evolution as previously believed 
because this process leads to merging of the components when giants have such
low-mass cores. Stable mass transfer in low-mass binaries, 
from more than 65,000 runs of Population I binary evolution, 
may well reproduce the properties of EL CVn stars such as the evolutionary phase, 
mass ratios and the WD mass-period ($M_{\rm WD}-P$) relation.  
The study shows that the most common donor mass range for producing the observed EL
CVn-type stars is $1.15-1.20M_\odot$ and that the lifetime of such objects
increases dramatically with decreasing proto-He WD mass.
This leads to an intrinsic peak mass around the minimum mass of proto-He WDs of $\sim 0.16M_\odot$. 
The most probable proto-He WD mass is $0.17-0.21M_\odot$ after selection effects included. 
We estimated the number of EL CVn stars with  $P\le2.2$ d to be $2-5\times
10^6$ in the Galaxy, indicating a local density of $4-10\times
10^{-6}{\rm pc}^{-3}$. We conclude that many more EL CVn-type
binaries remain to be discovered and that these binaries will be
predominantly low-mass systems in old stellar populations.
\end{abstract}




\section{Introduction}
Using the Wide Angle Search for Planets (WASP, Pollacco et al.
2006) photometric database, \citet{maxted14a} recently discovered
17 bright eclipsing binaries which contain an A- or F-type dwarf
(main sequence, MS) star and a helium white-dwarf precursor
(proto-He WD hereinafter), i.e. being in a rarely-observed state
evolving to higher effective temperatures at a nearly constant
luminosity prior to being a very low-mass WD.
Among these stars, EL CVn is the brightest and is the only one previously identified
as a variable star, so has been adopted as the prototype of this
class of eclipsing binary star. 
In this paper, we define a binary as an EL CVn-type binary if it has an A- or F-type dwarf
star (with mass of $1.05-2.9M_\odot$) and a proto-He WD being in the constant-$L$ phase.
Multi-periodic pulsations were
first discovered in the proto-He WD companion of 1SWASP
J024743.37-251549.2 (J0247-25 hereinafter) where the proto-He WD has
a mass of $\sim 0.2M_\odot$ \citep{maxted11, maxted13}, then in
1SWASP J162842.31+101416.7 (WASP 1628+10) where the mass of the
proto-He WD companion is only about
$0.135\pm0.02M_\odot$\citep{maxted14b}. This provides an
opportunity to study the structure and formation of such low-mass
WDs in details through the application of asteroseismology to this
new class of pulsating variable stars.

Objects similar  to the proto-He WD companions of EL CVn-type binaries,
i.e. low-mass, thermally bloated hot WDs have been discovered in the Kepler field. They are KOI 74, KOI
81\citep{van10}, KHWD3 \citep[KOI 1375,][]{carter11}, KOI 1224
\citep{breton12}, KIC 9164561 and KIC 10727668 \citep{rap15}. Some
of them are obviously the descendants of EL CVn- type stars --
they have just departed from the constant-$L$ phase and some may
be in the stage of surface H flash. Moreover, the RR Lyr star
OGLE-BLG-RRLYR-02792 was found to be an eclipsing binary system
where the pulsator is a proto-He WD with a mass of $0.26M_\odot$
\citep{nature12}. EL CVn-type stars are then probably common in
universe and it is necessary to study their formation in details
and estimate their number in the Galaxy. 

Extremely low-mass WD (ELM, with masses below $0.2-0.3M_\odot$, descendants of proto-He WDs) 
have also been found in some millisecond pulsars (MSPs, see a summary in Table 1 of \citet{ist14a}) 
and in double degenerate systems (DDs, see the latest paper of the ELM survey of \citet{brown16} ). 
In general, low-mass He WDs ($M_{\rm WD}\lesssim 0.4 M_\odot$) can be produced by binary evolution through stable mass transfer or rapid common-envelope evolution (CEE). The CEE is unlikely to produce the low-mass He WDs in MSPs since the neutron star should be spun up through stable mass accretion. 
Our simple analysis for the 17 EL CVn samples (see Appendix) show that  
EL CVn-type stars are also unlikely to be produced from the CEE, in which the donor (the progenitor
of the proto-He WD) should be at or very close to the base of red
giant branch (RGB) (due to the very low mass of the proto-He WDs, $\sim 0.2M_\odot$), 
where the envelope is tightly bound and the ejection of the CE results in a system with a much shorter orbital period e.g. some DDs in the ELM survey, than those of EL CVn stars.
We are therefore only concerned here with the stable
mass transfer channel for the formation of EL CVn-type stars.

Before we go into the details of stable mass transfer channel, 
we firstly introduce the bifurcation period, $P_{\rm b}$, in low mass binary
evolution, which is crucial for the formation of EL CVn-type
stars. Fig. 1 presents a group of evolutionary
tracks of the donors for binaries with component masses of 1.0 and
$0.9M_\odot$ but various initial orbital periods, $P_{\rm 0}$. We
can see two evolutionary regimes based on the value of $P_{\rm
0}$. The donor evolves to ever decreasing luminosity for short
$P_{\rm 0}$ while it passes through a nearly constant-$L$ phase
and leaves a degenerate remnant eventually when $P_{\rm 0}$ is
longer than the critical value $P_{\rm b}$. This is known as the
bifurcation period \citep[see also][]{nelson04}. Binaries below
the bifurcation period ($P_{\rm 0}< P_{\rm b}$) typically evolve
to a minimum orbital period (on the order of 1 hr) and then back
to slightly increasing orbital periods. For binaries with $P_{\rm
0}> P_{\rm b}$, there is a fairly tight $M_{\rm WD}-P$ relation,
where $M_{\rm WD}$ and $P$ are the remnant mass of the donor and
the final orbital period, respectively. The $M_{\rm WD}-P$
relation has been observed in several types of close binary stars
such as binary radio pulsars \citep{rap95,lin11,smed14,tauris14},
hot subdwarfs \citep{chen13} and a number of Kepler binaries
containing low-mass hot WDs \citep{rap15}. The predicted value of
$P_{\rm b}$ depends on the initial parameters of each binary and
the assumptions in the modelling of the binary evolution process.

\begin{figure}
\includegraphics[width=6.0cm,angle=270]{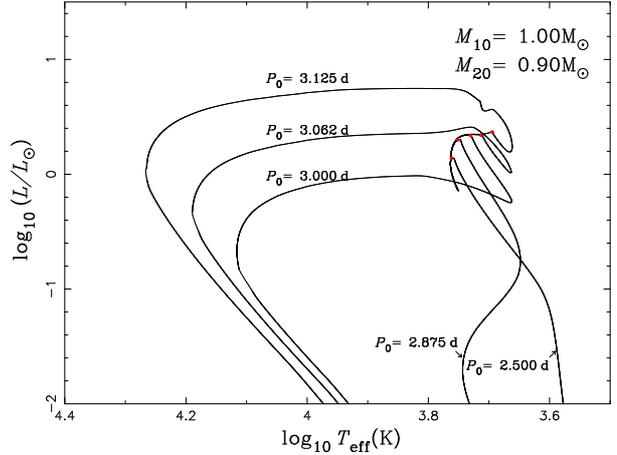}
\caption{Evolutionary tracks on the Hertzsprung-Russel diagram for
binaries with the same initial component masses ($M_{\rm 10}$, $
M_{\rm 20}$) but various initial orbital periods ($P_{\rm 0}$), as
denoted in the figure. It clearly shows the bifurcation
period in binary evolution: for initial orbital period $P_{\rm 0}$
less than the bifurcation period $P_{\rm b}$ ($P_{\rm b}\simeq3$ d
in this case) the mass donor evolves directly to very low
luminosity, and for $P_{\rm 0}$ slightly greater than $P_{\rm b}$
the evolution can produce a long-lived phase of evolution at
nearly constant luminosity (proto-He WD phase).  \label{bifu}}
\end{figure}

The very low mass of the proto-He WDs and the short orbital periods
mean that only binaries with initial orbital periods slightly
longer than $P_{\rm b}$ can produce EL CVn-type stars. For
example, the reconstructions of the evolutionary history for WASP
J0247-25 and WASP 1628+10 \citep{maxted13, maxted14b} showed that
both the objects are probably from stable mass transfer where the
mass donor starts to lose mass before it leaves the MS, i.e. the
central H abundance is larger than zero. The parameter space for
producing EL CVn-type stars in this way is then expected to be
small since both $M_{\rm WD}$ and $P$ increase significantly with
$P_{\rm 0}$. The expectation is against the fact that  many such
objects have been discovered. The timescale for the nearly
constant-$L$ phase, $t$, may play a role here. Some particular
examples \citep[e.g.][]{nelson04} show that the value of $t$ can
be up to nearly $10^9$ yrs when the proto-He WD mass is lower than
about $0.21M_\odot$. The majority of the products in
\citet{nelson04} have He WDs larger than $0.2 M_\odot$, more
massive than those in EL CVn-type stars. 
\citet{ist14a, ist14b} studied the formation of the low-mass He WD in MSPs 
through stable mass transfer near the bifurcation period and 
obtained a very strong dependence of the proto-He WD lifetime on its mass,
approximately as the reciprocal of WD mass  to the 7th power. 
And very recently, they \citep{ist16} give the detailed models of the low-mass He WD produced in low-mass X-ray binaries including some microphysics such as gravitational settling, thermal and chemical diffusion, and rotational mixing. In principle, the proto-He WDs in EL CVn-type binaries 
have similar properties to those in MSPs.  However, 
a complete binary evolution study and a binary population synthesis are still necessary
for our comprehensive understanding of the formation and
characteristics of EL CVn-type stars, e.g which binary systems can
produce EL CVn-type stars, whether the number of such binaries
from the stable mass transfer channel is sufficient to explain the
observations, whether the binary properties are consistent with
observations, what are the minimum mass and the peak mass, etc.

We carried out more than 65,000 runs of binary evolution for
Population I (Pop I) low-mass binaries for this comprehensive
study. The details for the computation are introduced in Section
2, and the evolutionary consequences are summarized in Section 3.
In Section 4, we show the number and parameter distributions of EL
CVn-type stars from our binary population synthesis. In Section 5,
we give some discussions, and the conclusions are shown in Section
6.

\section{The Computations}
We carried out binary evolution calculations with a
newly-developed Henyey code called MESA \citep[Modules for
Experiments in Stellar Astrophysics,][]{bill11,bill13,bill15}. The
advantages of MESA for our calculations are the inclusion of
equations of state that are able to handle the low mass and high
degree of degeneracy that the donors achieve, and the powerful
capabilities of handling the huge fluctuations during the
evolution such as H or He flashes.

We used MESA version 3635, in which the mass transfer scheme is
similar to that of the Ritter scheme in the latest version of MESA
\citep{bill15}, that is, if we denote the density by $\rho_{\rm
L1}$,the isothermal sound speed by $v_{\rm s}$, and the effective
cross section of the flow by $Q$ (all taken at $L_{\rm 1}$), the
mass transfer rate is given by \citep{ritter88}

\begin{equation}
\dot M_{\rm RLOF}=-\rho _{\rm L1}v_{\rm s} Q\propto{\rm
exp}(\frac{R_{\rm 1}-R_{\rm L1}}{H_{\rm p}}),
\end{equation}
where $R_{\rm 1}$ and $R_{\rm L1}$ are the radii of the donor and
its Roche lobe, respectively, and $H_{\rm p}$ is the pressure
scale height of the atmosphere divided by a factor related to the
mass ratio, as described in \citet{ritter88}. This expression
allows mass transfer to take place through the inner Lagrangian
point $L_{\rm 1}$ even when $R_{\rm 1}<R_{\rm L1}$, which is
reasonable since stars generally have extended atmospheres
\footnote{In our calculation, we used the default option for the atmosphere boundary condition, that is, `$simple_{\rm -}photosphere$' , an Eddington approximation estimated for optical depth equal to 2/3. The choice of this bounary condtion almost has no effect on our results in this paper since we are only concerned on the temperature but not the spectral energy distribution.}. 

The mass transfer is assumed to be non-conservative, with 50 per
cent of the material lost from the donor being accreted by the
companion while the other 50 per cent leaves the system with the
same specific angular momentum as pertains to the donor. 
The treatment of non-conservative mass transfer, 
in particular the amount of specific angular momentum that is lost from the system
is a major uncertainty in modeling binary evolution and  
different reasonable prescriptions can give very different evolutionary paths \citep{philipp08,egg10,lin11,ist14a}.
Varying the assumptions on the amount of mass and angular momentum loss will affect the value of the bifurcation period, resulting in a change in the parameter space available for the
production of EL CVn-type stars, but has little effect on the
properties such as the lifetime of the proto-He WDs and the proto-HeWD mass-orbital period relation etc. We will give a discussion on this in Sects. 3 and 4.

We have not followed the evolution of the companions since it is very time consuming. For convenience, we simply assume that the companion stars are in thermal equilibrium and obtain their parameters such as radius $R$, luminosity $L$ from the classical relations for main sequence stars, depending on their masses $M$, that is \citep{liyan14},

\begin{equation}
{\rm log} (R/R_\odot)=\left\{
 \begin{array}{lc}
 0.73{\rm log}(M/M_\odot), M >0.4M_\odot,\\
 {\rm log}(M/M_\odot)+0.10, M<0.4M_\odot\\
\end{array}\right.
\end{equation}

\begin{equation}
{\rm log} (L/L_\odot)=\left\{
 \begin{array}{lc}
 4.0{\rm log}(M/M_\odot)+0.0792, M >M_\odot,\\
 2.76{\rm log}(M/M_\odot)-0.174, M<M_\odot.\\
\end{array}\right.
\end{equation}

The behavior of the accreting companion is complicated, e.g. it will lose thermal equilibrium when the mass accreting rate is larger than a certain value (depending on the mass of the accretor) and expand \citep{kipp77}, and may overfill its Roche lobe, leading to the formation of a contact binary and the subsequent termination of the formation process of EL CVn stars \citep{nel01}. 
We stop the calculations when the accretor overfills its Roche lobe even being in thermal equilibrium 
(i.e. the radius is obtained from Eq(2)) or the luminosity of the donor drops below ${\rm log}(L/L_\odot)\le-2.5$ (the system evolves below the bifurcation period and the donor cannot contract to be a He WD) and discard the models if the mass transfer rate $\dot{M}$ is larger than $10^{-4} M_\odot{\rm /yr}$ since (1) $\dot{M}$ will be up to 1$M_\odot$/yr soon after $10^{-4}M_\odot{\rm /yr}$ in most binaries that we checked (see section 3.1), leading to dynamically unstable mass transfer\footnote{Extensive effort has been put into developing reliable criteria to
distinguish between dynamically stable and unstable mass transfer
\citep{hjellming87,webbink88,soberman97,han02,chen08,wood12,pav15,ge15},
but there are still many problems and no consensus on this issue. }, and (2) the accretor will expand rapidly within such a high accretion rate and overfill its Roche lobe very likely (the binary cannot evolve to an EL CVn binary in both cases). 

The main sequence lifetime of the companion (from the zero-age main sequence to the termination of main sequence), $t_{\rm MS}$, is obtained from the fitting formula in Hurley's paper \citep{hurley02}. Since the companion has evolved for a while before mass accretion, the companion star is not exactly at the zero-age main sequence and the $t_{\rm MS}$ here is in fact an upper limit. But we checked that it has little effect on the final result. We reduced the main sequence lifetime of the companion to be 0.5$t_{\rm MS}$ and obtained a very similar result as shown in section 3.

Gravitational wave radiation (GWR) is accounted for using the
following equation (Landau \& Lifshitz, 1971):

\begin{equation}
\dot J_{\rm GR}=-\frac{32}{5} \frac{G^{7/2}}{c^5}\frac{M_{\rm
1}^2M_{\rm 2}^2(M_{\rm 1}+M_{\rm 2})^{1/2}}{a^{7/2}},
\end{equation}
where $M_{\rm 1}$, $M_{\rm 2}$ are component masses of the binary,
$a$ is the semi-major axis of the (circular) orbit, $c$ is the
speed of light in vacuum and $G$ is the gravitational constant. To
account for magnetic braking (MB) we use equation (36) from
Rappaport et al. (1983) with $\gamma=3$, i.e.,
\begin{equation}
\dot J_{\rm MB}=-3.8\times 10^{-30}M_{\rm 1}R_{\rm 1}^3\omega^3
{\rm dyn\cdot cm},
\end{equation}
where $M_{\rm 1}$ and $R_{\rm 1}$ are the mass and radius of the
donor and $\omega$ is its rotational angular velocity which is
assumed to be equal to the orbital angular velocity of the system
due to tidal synchronization. We switch off the MB when the
surface convective region is too small, i.e. less than 1 per cent
of the total mass.

Our study is focused on Pop I low-mass binaries (with a
metallicity $Z=0.02$) and the initial parameters of our models are
summarized as follows:

The donor mass ($M_{\rm 10}$ in $M_\odot$) : 0.90, 0.95, 1.0,
1.05, 1.10,..., 2.0 (with $\Delta M=0.05M_\odot$)

Initial mass ratio ($q_{\rm 0}=M_{\rm 10}/M_{\rm 20}$, where
$M_{\rm 20}$ is the initial mass of the secondary) : 1.1, 1.2,
1.3, ..., 4.0 (with $\Delta q_{\rm 0}=0.1$)

Initial orbital period ($P_{\rm 0}$ in days, by step of $\Delta
P_{\rm 0}=0.02$): from 2.5 to 4.5 for $M_{\rm 10}=0.9-1.05M_\odot$
and 0.5 to 2.5 for $M_{\rm 10}\ge 1.10M_\odot$.

Various values of $P_{\rm 0}$ for different donor masses and such
a small interval in $P_{\rm 0}$ are chosen so that we obtain a
relatively accurate value of $P_{\rm b}$ and a dense grid around
$P_{\rm b}$ for each set of ($M_{\rm 10}$, $q_{\rm 0}$). This is
important since the parameter space for EL CVn-type stars is
expected to be {\it just beyond} $P_{\rm b}$, as mentioned in
Section 1.

\section{Binary Evolution Results}
\subsection{Parameter Space Leading to the Formation of EL CVn Binaries}
\begin{figure}
\includegraphics[width=6.0cm,angle=270]{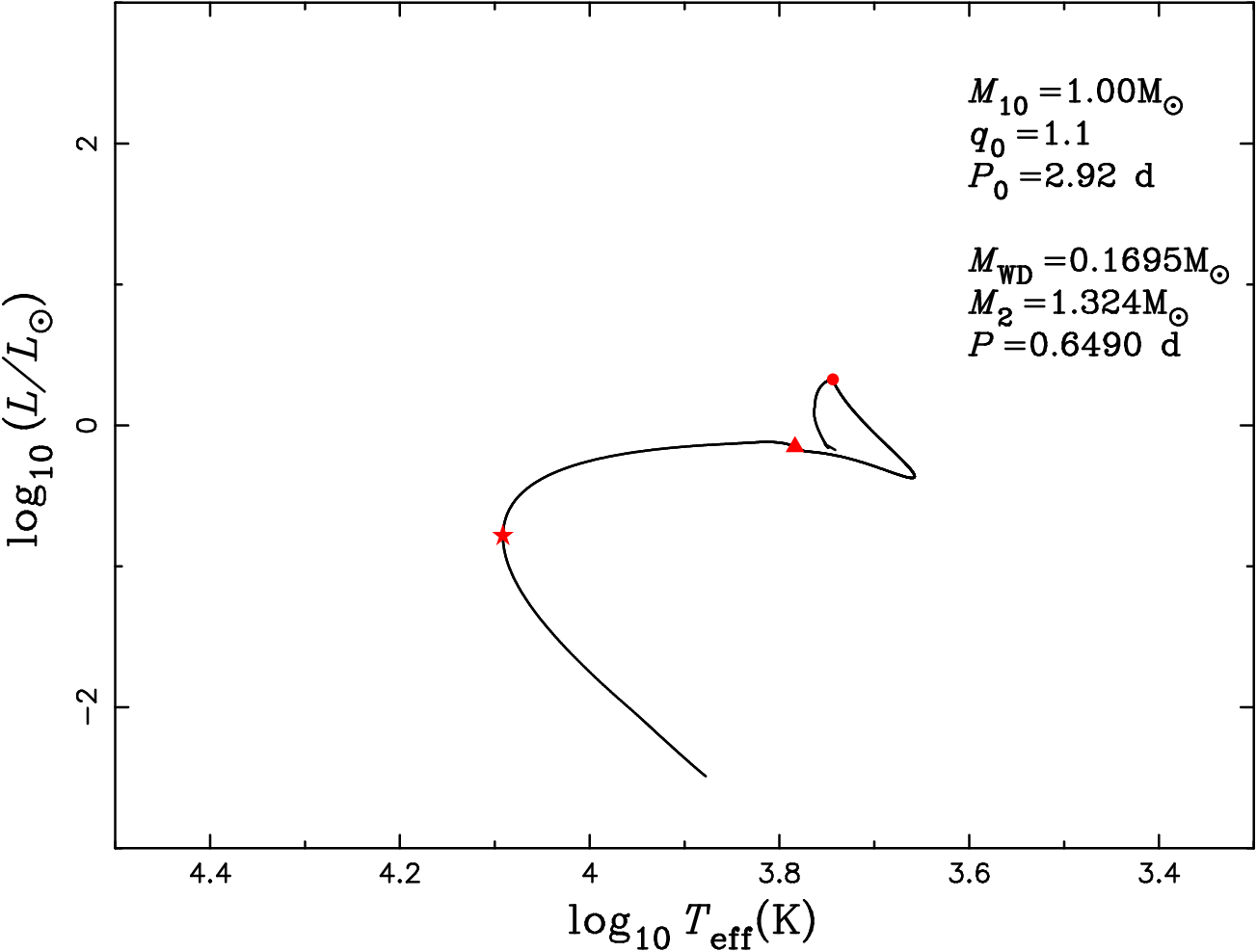}
\includegraphics[width=6.0cm,angle=270]{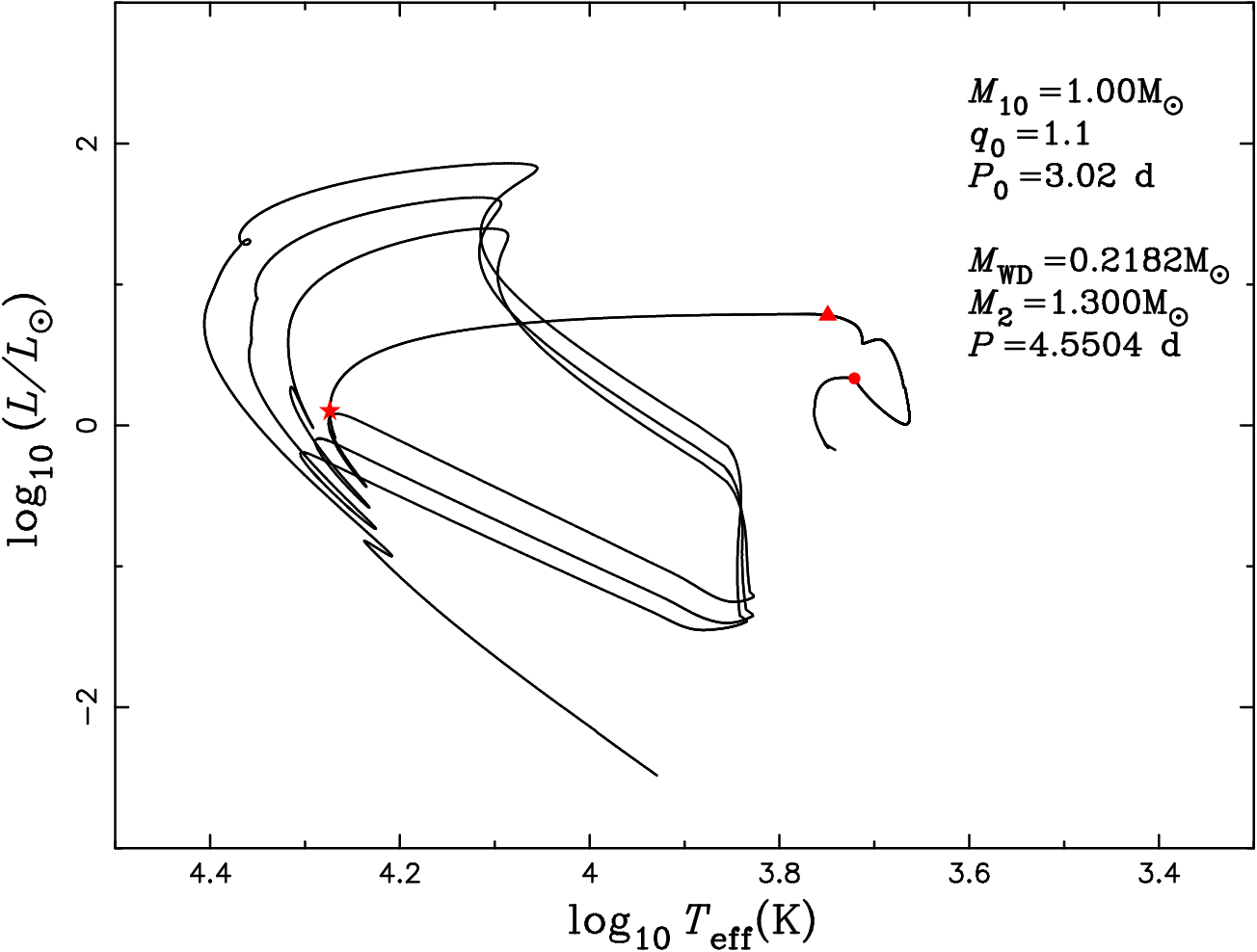}
\caption{Two typical evolutionary tracks in our study on the
Hertzsprung-Russel diagram. The binaries have the same component
masses, that is, $M_{\rm 10}=1.00M_\odot$ and $q_{\rm 0}=1.1$. The
initial period and final parameters for each binary are indicated
in the figure. The circles, triangles and stars indicate the onset
of mass transfer, the end of mass transfer and the end of the
nearly constant luminosity phase, respectively. For the binary in
the upper panel, $P_{\rm 0}>P_{\rm b}$ but very close to $P_{\rm
b}$, the donor does not ascend the RGB during mass transfer and
cools directly to become a white dwarf after the nearly
constant-$L$ phase. For the one in the lower panel, $P_{\rm 0}$ is
a little bit larger than $P_{\rm b}$, and the donor ascends the
RGB during mass transfer. The remnant suffers several H shell
flashes after the nearly constant-$L$ phase before it
cools down.\label{hrd}}
\end{figure}

Two typical evolutionary tracks for EL CVn-type stars in our
calculation are presented in Fig. 2, where the two binaries have
the same component masses, $(M_{\rm 10},q_{\rm 0})=(1.00,1.1)$,
but different initial periods, that is, $P_{\rm 0}=2.92$ d for the
upper panel and $P_{\rm 0}=3.02$ d for the lower one. For the
binary in the upper panel, $P_{\rm 0}>P_{\rm b}$ but very close to
$P_{\rm b}$, the donor does not ascend the RGB during mass
transfer and the core has not been degenerate yet at the
termination of mass transfer (defined as ${\dot M}_{\rm RLOF}\le
10^{-12}M_\odot {\rm yr}^{-1}$). The remnant has a final mass of
$M_{\rm WD}=0.17M_\odot$ and cools directly to become a WD after
the nearly constant-$L$ phase. For the binary in the lower panel,
however, $P_{\rm 0}$ is a little bit larger than $P_{\rm b}$, the
donor ascends the RGB during mass transfer and the core is
degenerate at the end of mass transfer. As a consequence, the
remnant has a larger final mass ($M_{\rm WD}=0.22M_\odot$) and
suffers several H shell flashes after the nearly constant-$L$
phase but before the cooling. The H shell flashes cause the
luminosity of the remnant to increase by several orders of
magnitude on a dynamical timescale. The remnant is then highly
bloated during the flashes and has characteristics very different
from that in the constant-$L$ phase. In this paper, we only focus
on the nearly constant-$L$ phase before H flashes, and define the
end of the nearly constant-$L$ phase to be the point of maximum
$T_{\rm eff}$ before H flashes or the He WD cooling if no flash
occurs.

\begin{figure*}
\includegraphics[width=9.0cm,angle=270]{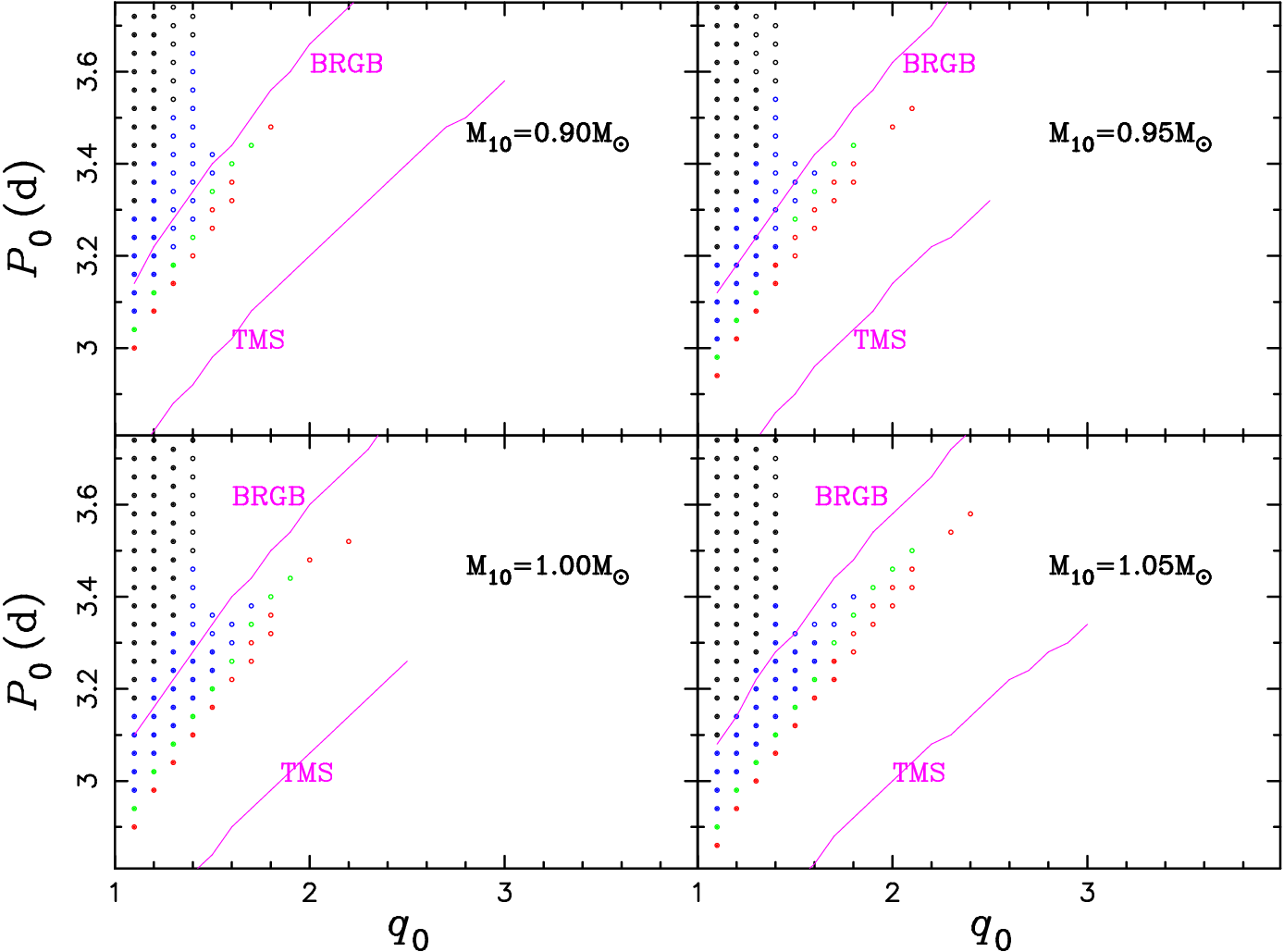}
\includegraphics[width=9.0cm,angle=270]{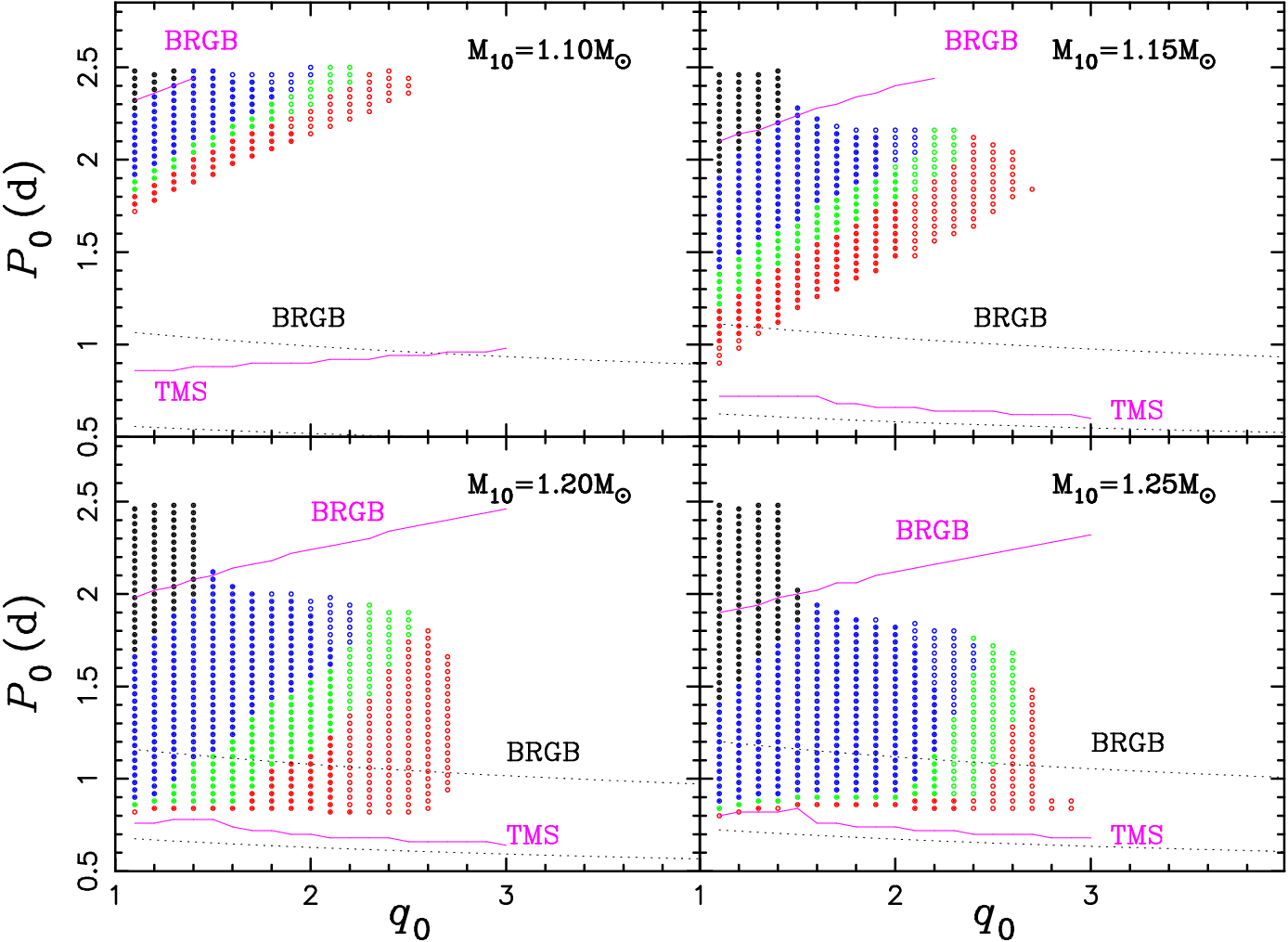}
\caption{Summary of our calculations in the ($P_{\rm 0}, q_{\rm
0}$) plane. Symbols of open circles indicate the binaries which can produce He WDs 
and the filled circles overlapped on the open circles are those which can produce EL CVn-type stars.
The initial donor mass is indicated in each panel.
Various orbital periods at the end of mass transfer are denoted
with different colors, i.e. the red, green, blue and black are for
$P$ in the range of 0.5--1.0 d, 1.0--2.2 d, 2.2--10 d and $>10$
days, respectively. Models not presented are those binaries that
evolve into contact, below the bifurcation point, and ${\dot M}_{\rm RLOF}>
10^{\rm -4}M_\odot {\rm yr}^{-1}$.  
The purple solid lines show the boundaries for models which starts mass transfer at the
termination of MS (TMS) and at the base of RGB (BRGB), while the
black dotted ones are for the same boundaries for models where
magnetic braking is not included (The dotted lines are below the minimum value of the respective $y$-axis variables in the upper four panels and are not presented). For clarity, one model was
removed between every two points. \label{grid}}
\end{figure*}

\begin{figure*}
\includegraphics[width=9.0cm,angle=270]{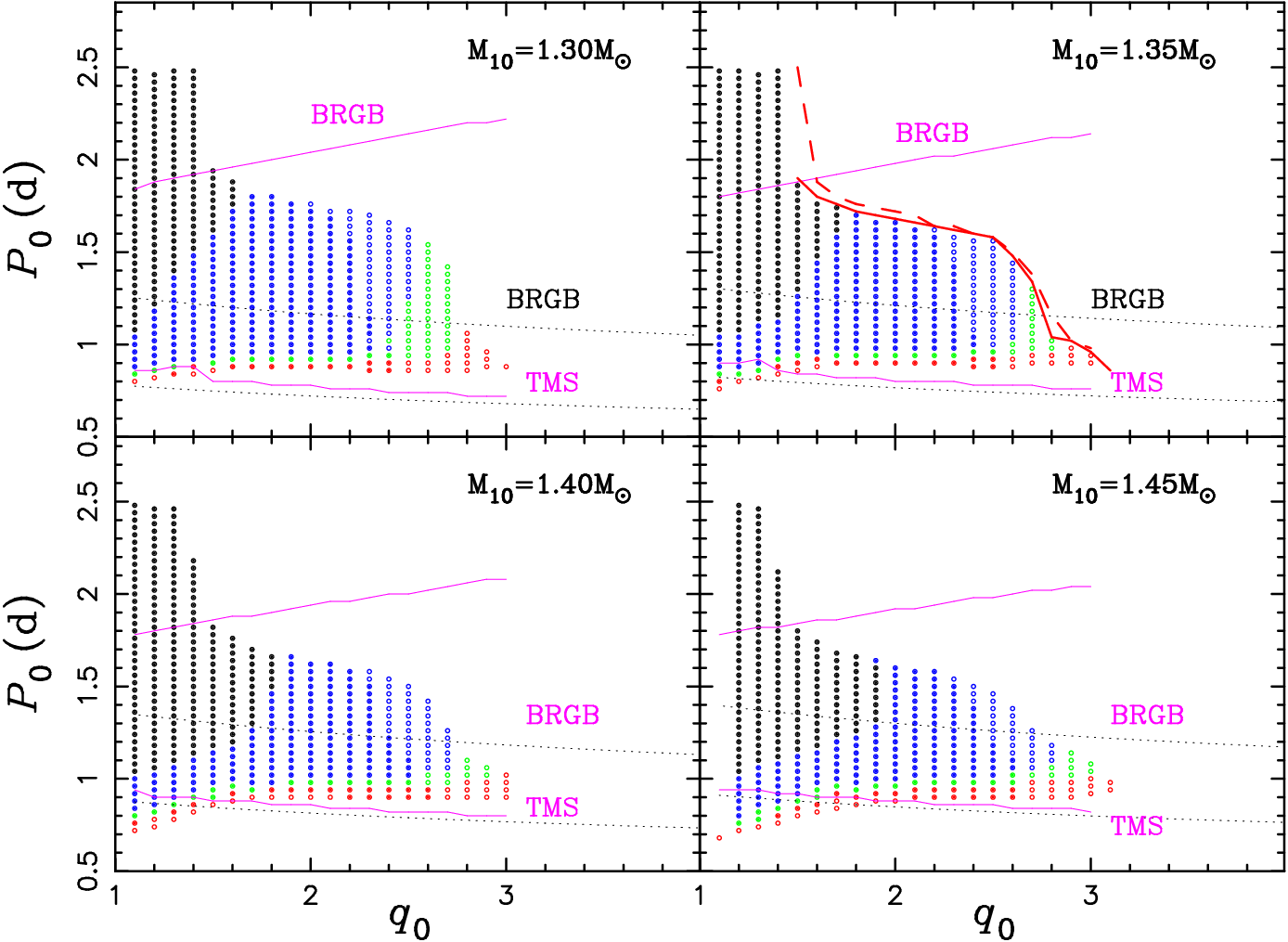}
\includegraphics[width=9.0cm,angle=270]{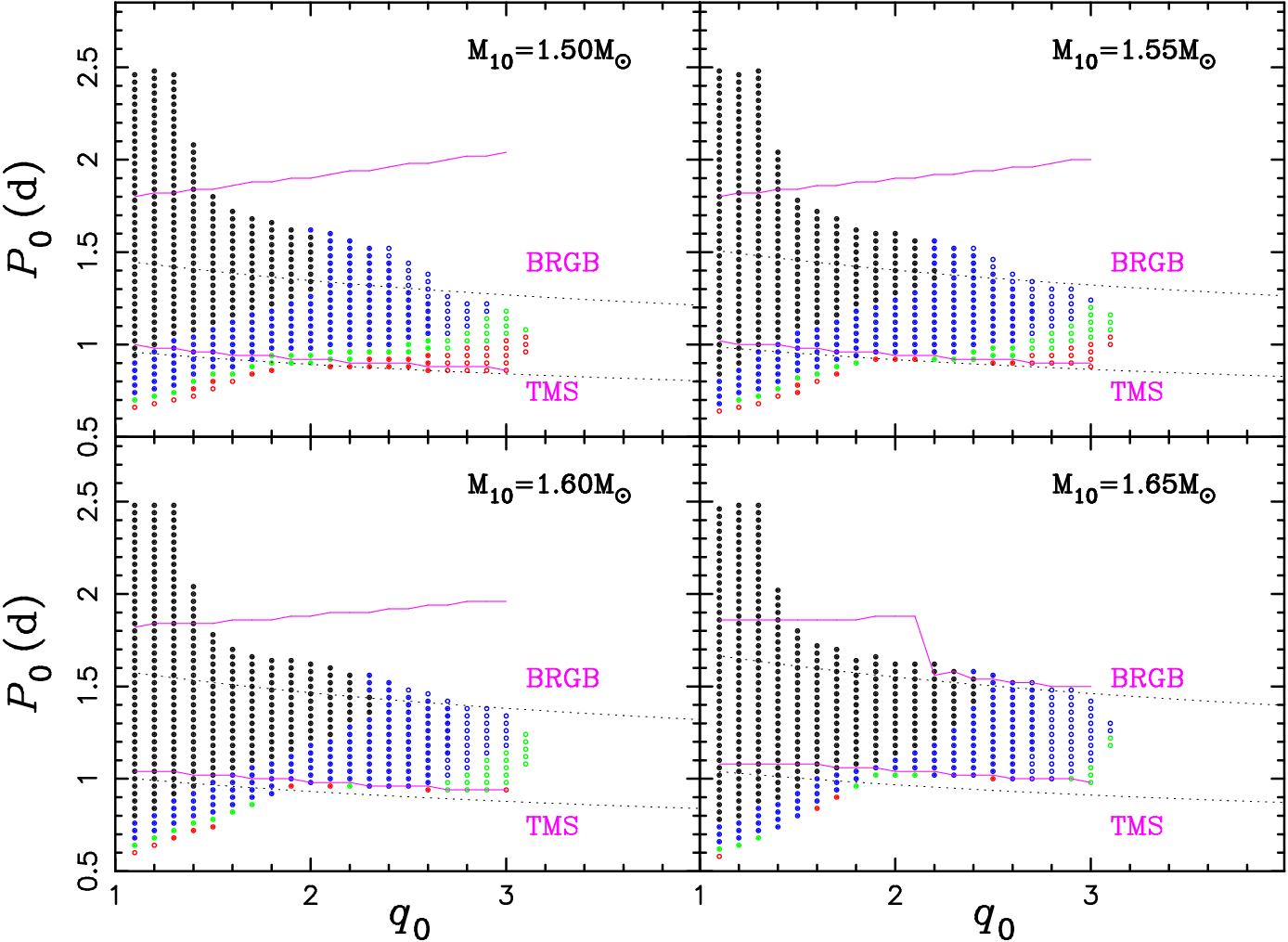}
\caption{Similar to Fig.3 but for $M_{\rm 10}=1.30-1.65M_\odot$. The thick solid and dashed lines in the upper right panel ($M_{\rm 10}=1.35M_\odot$) show the boundaries for mass transfer rate of $10^{\rm -4}M_\odot {\rm yr}^{-1}$ and $1M_\odot {\rm yr}^{-1}$, respectively.
\label{grid-2}}
\end{figure*}

\begin{figure*}
\includegraphics[width=9.0cm,angle=270]{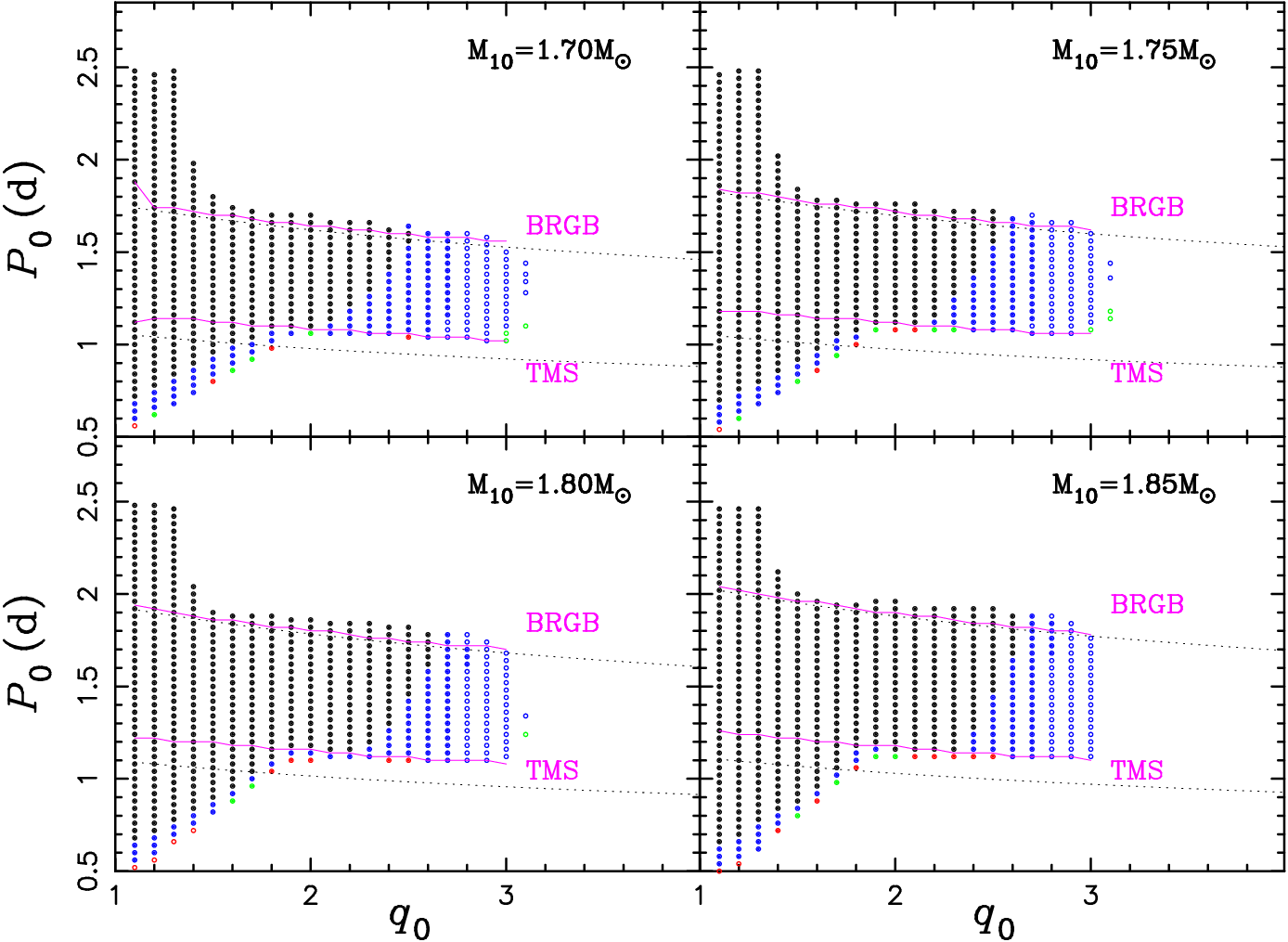}
\includegraphics[width=9.0cm,angle=270]{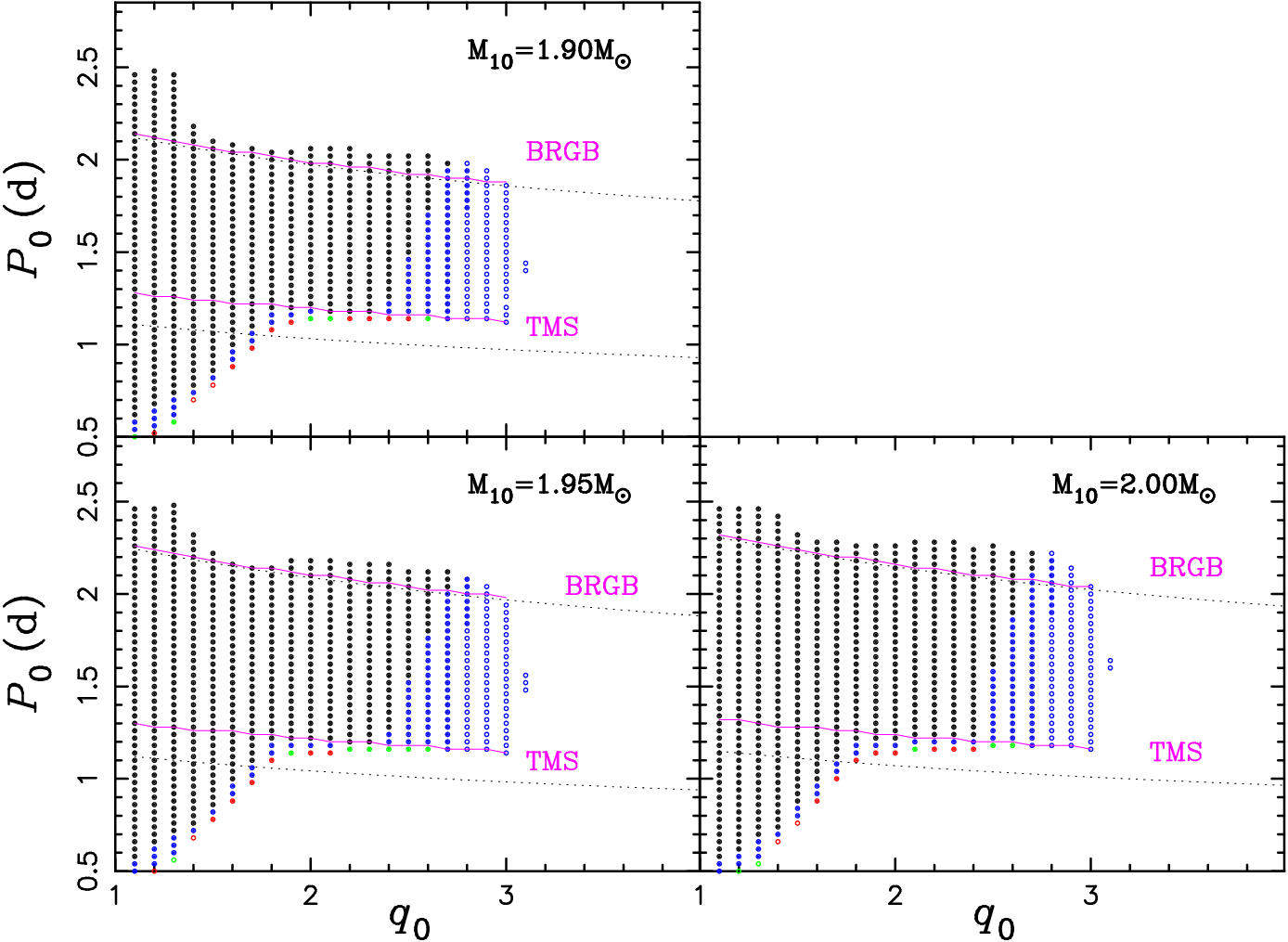}
\caption{Similar to Fig.3 but for $M_{\rm
10}=1.70-2.00M_\odot$.\label{grid-3}}
\end{figure*}

The evolution paths predicted by our calculations are summarized
in Figs. 3--5. According to observations, EL CVn-type binaries are
defined as the binaries consisting of a proto-He WD and an A- or F-
type dwarf companions, i.e an MS star with its mass in the range
of $1.05-2.9M_\odot$. In each panel of the figures, the lower
boundary is determined by the bifurcation period, while the upper and right boundaries are
determined by the limit of maximum mass tranfer rate ($10^{\rm -4}M_\odot {\rm yr}^{-1}$).
From the figures we see that when ${M_{\rm 10}<}1.30M_\odot$ only binaries in which the donor
starts mass transfer after the end of MS can leave degenerate
remnants and so contribute to the production of EL CVn-type stars.
For more massive donors EL CVn-type stars can be produced when the
donor starts mass transfer on the MS if $q_{\rm 0}$ is not very
large. The most likely mass for producing short orbital period EL CVn-type stars (i.e. $P\le2.2$ d) is in the range of $1.15-1.20M_\odot$. For $M_{\rm 10}>1.2M_\odot$, the parameter space decreases with the mass increasing and few models can produce EL CVn-type binaries with $P<2.2$ d when $M_{\rm 10}\ge1.60M_\odot$. We see also small parameter spaces for $M_{\rm 10}\le1.10M_\odot$. We explain this as follows.

It is the MB that determines the parameter space for producing EL CVn stars. The discontinuous change between $1.10M_\odot$ and $1.15M_\odot$ is due to the fact that MB works when the donor is on the MS for $M_{\rm 10}\le1.10M_\odot$ but does not for $M_{\rm 10}\ge1.15M_\odot$. When $M_{\rm 10}\le1.10M_\odot$,  many binaries that would evolve beyond the bifurcation period (then producing proto-He WDs) now evolve below the bifurcation period since the MB leads the mass transfer starting in advance. These binaries then cannot produce EL CVn stars any more, resulting in the small parameter space in low-mass range. 
For binaries with $M_{\rm 10}\ge1.15M_\odot$, the convection envelope develops in the late phase of Hertzsprung gap (HG) and the MB plays a crucial role in that phase, i.e.  many of the progenitors of the proto-He WDs, which would start mass transfer on RGB if MB were not included, start mass transfer in advance, resulting in stable mass transfer and low-mass He WDs are more likely. With the increasing of the donor mass, the mass transfer rate becomes larger and larger in HG and may be up to $10^{\rm -4}M_\odot {\rm yr}^{-1}$ before the donor develops the surface convection region, and the MB does not affect the result any more. This occurs when $M_{\rm 10}>1.65M_\odot$. To trace the effect of MB on parameter space, we show the boundaries at which the donor overfills its Roche lobe at the termination of the MS (TMS) and at the base of RGB (BRGB) when the MB has been included (the thin solid lines) and when the MB  has not been included (the dotted lines) in Figs. 3-5. We see that the TMS lines in the two cases are similar when $M_{\rm 10} \ge 1.15M_\odot$ and the BRGB lines are similar when $M_{\rm 10} > 1.65M_\odot$, clearly showing the parameter space affected by the MB, as we described above. 
Note that, even when the MB does not play a role with mass increasing gradually,  the TMS (BRGB) lines in our models are not exactly overlapped with that of no MB effect included since we adopted Eq(1) to calculate the mass transfer rate; that is, the mass transfer starts before the donor overfills its Roche lobe.

As mentioned before, we have not included the models with mass transfer rates exceeding $10^{\rm -4}M_\odot {\rm yr}^{-1}$  in Figs. 3-5. However it is very valuable for binary evolution to trace the following behaviours of the donor when the mass transfer rate is larger than $10^{\rm -4}M_\odot {\rm yr}^{-1}$. 
We then set the maximum allowed mass transfer rate to be 1$M_\odot{\rm yr}^{-1}$ in our calculation and recalculate the models of $M_{\rm 10}=1.35M_\odot$.
In our original parameter space, the mass transfer rates for binaries with this donor mass are always smaller than $10^{\rm -4}M_\odot {\rm yr}^{-1}$ when the initial mass ratio $q_{\rm 0} \le 1.4$. We therefore only examined the models with $q_{\rm 0} \ge 1.5$.  We found that, in a few models, the mass transfer rate may exceed $10^{\rm -4}M_\odot {\rm yr}^{-1}$ but does not increase up to $1M_\odot {\rm yr}^{-1}$. 
However, in most binaries we studied, the mass transfer rate will increase up to 1$M_\odot{\rm yr}^{-1}$
 soon after $10^{\rm -4}M_\odot {\rm yr}^{-1}$ is reached. 
 The timescale of this process (from $10^{\rm -4}M_\odot {\rm yr}^{-1}$ to $1M_\odot{\rm yr}^{-1}$) 
is a few years to $\sim 300$ yrs, 
and the amount of mass transferred during this process is less than $0.1M_\odot$.
For example, if $q_{\rm 0}=1.7$, the mass transfer rate is always less than $10^{\rm -4}M_\odot {\rm yr}^{-1}$ 
when  $P_{\rm 0} \le 1.76$ d,  and it exceeds $10^{\rm -4}M_\odot {\rm yr}^{-1}$ for a while but never reaches $1M_\odot {\rm yr}^{-1}$ when $P_{\rm 0} = 1.78$ and 1.80 d. When $P_{\rm 0} \ge 1.82$ d, the mass transfer rate increases up to 1$M_\odot{\rm yr}^{-1}$ soon after $10^{\rm -4}M_\odot {\rm yr}^{-1}$ is reached, with a timescale less than 300 yr and the amount of mass transferred less than $0.1M_\odot$.
The dashed line in the upper right panel of Fig. 4 shows the boundary for $1M_\odot {\rm yr}^{-1}$, very close to that for $10^{\rm -4}M_\odot {\rm yr}^{-1}$. 
This indicates that the mass transfer rate of $10^{\rm -4}M_\odot {\rm yr}^{-1}$ might be a good (though not very strict) criterion for dynamical instability. 
Please note that the models with initial orbital periods in the gap between the boundary for $10^{\rm -4}M_\odot {\rm yr}^{-1}$ and that for  $1M_\odot {\rm yr}^{-1}$ have no contribution to the formation of EL CVn stars, though the mass transfer process is probably dynamically stable, since the companions will lose thermal equilibrium in such high mass accretion rates and expand to overfill their Roche lobes. The systems then become contact and the formation process of EL CVn stars is terminated (see section 2).   

\subsection{Lifetime during the proto-He WD phase}

\begin{figure}
\includegraphics[width=6.0cm,angle=270]{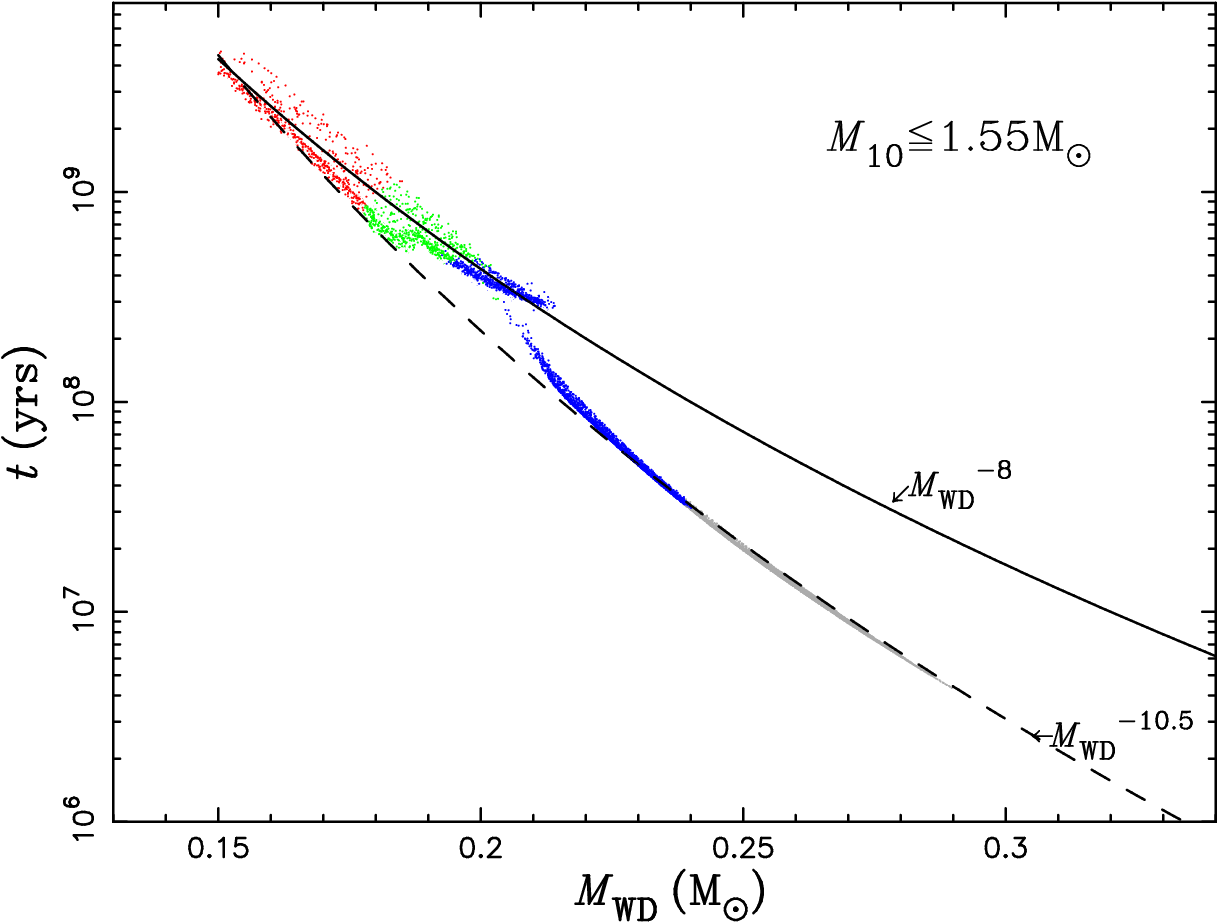}
\includegraphics[width=6.0cm,angle=270]{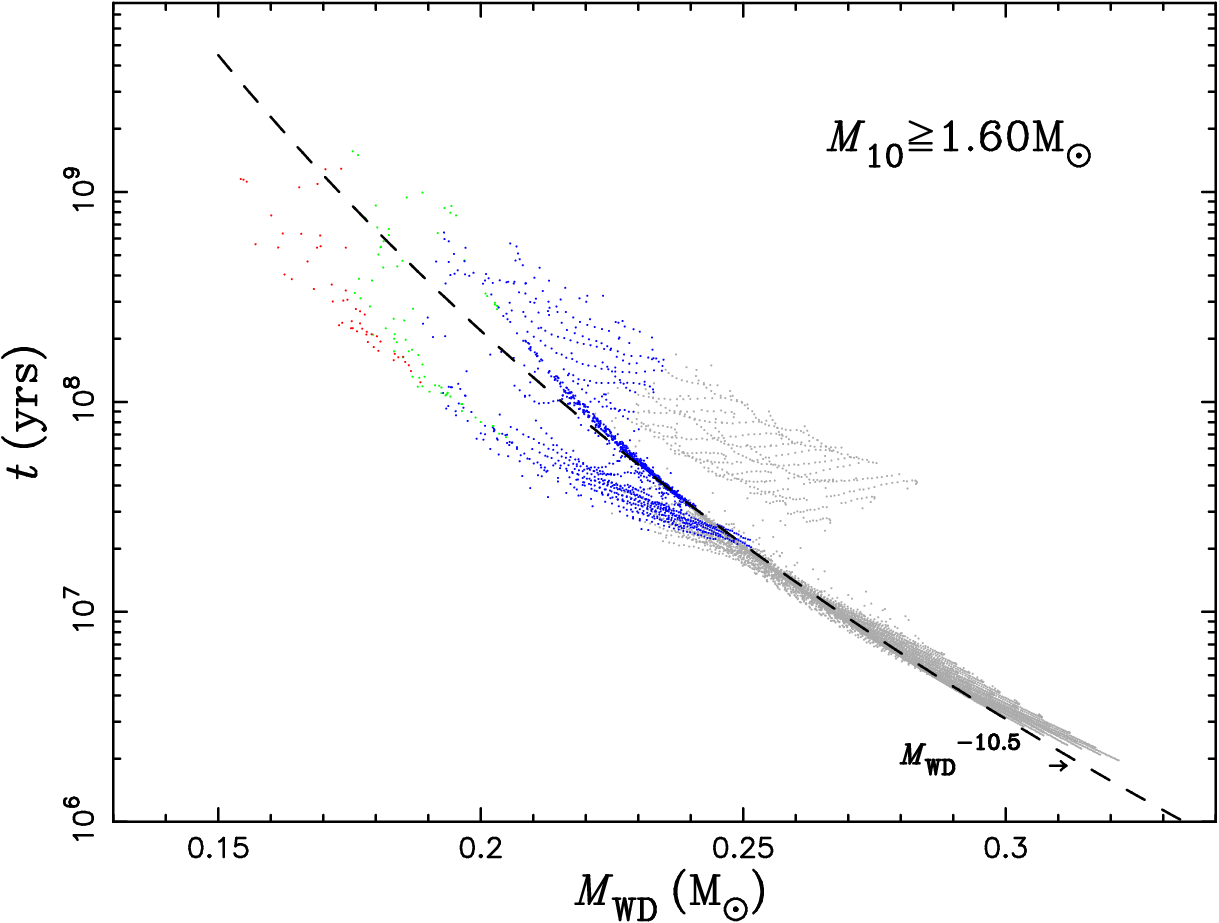}
\caption{Timescale for the nearly constant luminosity phase versus
the final mass of the donor, $M_{\rm WD}$, for EL CVn-type stars.
The colors of red, green, blue and gray are for the final orbital
period in the range of 0.5--1.0 d, 1.0--2.2 d, 2.2--10 d and $>10$
d, respectively. The upper panel is for $M_{\rm 10}\le1.55M_\odot$
and the lower panel for $M_{\rm 10}\ge 1.60M_\odot$. The solid and
dashed lines show two exponential relationships between $t$ and
$M_{\rm WD}$ (Eqs.(6) and (7) in the text) as indicated in the figure. \label{mftc}}
\end{figure}

\begin{figure}
\includegraphics[width=6.0cm,angle=270]{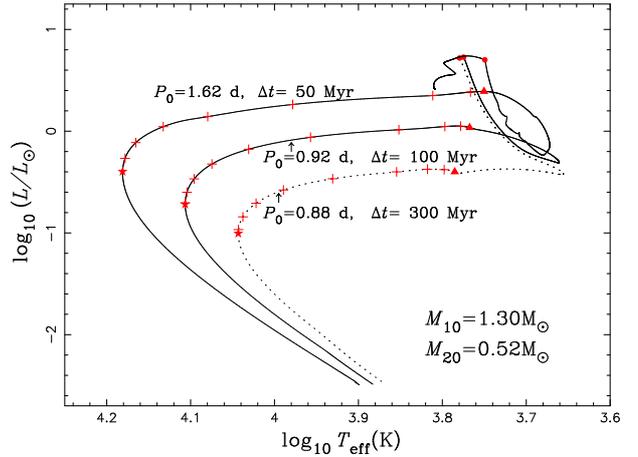}
\caption{Evoultionary tracks of the donors from the zero age main sequence to a WD 
and the time elapsed during the proto-He WD phase. Initial parameters of each bianry are indicated in the figure.
The circles, triangles and stars are for the onset of mass transfer, the end of mass transfer, and the end of
constant luminosity phase, respectively. The age differences between adjacent crosses, $\Delta t$, are equal and
have been indicated in the figure for each track.
 \label{tp}}
\end{figure}

Figure 6 shows the dependence of the lifetime in the nearly
constant-$L$ phase, $t$, on the proto-He WD mass, and Fig.7 presents three examples for the time elapsed during the proto-He WD phase. In Fig.6, the models are
separated into two populations according to whether the core of
the donor is degenerate or not. Similar to Figs. 3--5, various
colors denote different orbital period ranges. For models with
donors having degenerate cores (the upper panel), we see a strong
correlation between $M_{\rm WD}$ and $t$, that is, 
\begin{equation}
t  = 1100M_{\rm WD}^{-8}
\end{equation}
for products with $M_{\rm WD}\le0.21M_\odot$, and 
\begin{equation}
t = 10M_{\rm WD}^{-10.5}
\end{equation} 
for the lower lifetime limit. As mentioned in section 1,
 \citet{ist14b} obtained a similar strong dependence 
 (approximately as the reciprocal of WD mass to the 7th power)
between the lifetime of the proto-He WD and its mass according to their studies on MSPs. 
The strong correlation can be well explained as follows. After the
termination of mass transfer, the H-burning shell still exists and
supports the luminosity. The value of $t$ then can be simply
estimated by $M_{\rm env}/\dot{M}_{\rm nuc}$, where $M_{\rm env}$
is the envelope mass of the proto-HeWD, and $\dot{M}_{\rm nuc}$ is
the nuclear reaction rate of H and depends on the temperature in
the H-burning shell, $T_{\rm b}$. In general, $T_{\rm b}$
increases with core mass, $M_{\rm c}$, and $M_{\rm c} \simeq
M_{\rm WD}$ at the termination of mass transfer for donors with
degenerate cores. A small $M_{\rm WD}$ therefore indicates a
smaller $M_{\rm c}$, then a lower $T_{\rm b}$ and a lower
$\dot{M}_{\rm nuc}$. On the other hand, there is an
anti-correlation between $M_{\rm env}$ and $M_{\rm WD}$ (Fig. 8),
that is, a small $M_{\rm WD}$ has a larger $M_{\rm env}$ at the
end of mass transfer. Both of these factors result in a longer
lifetime for lower $M_{\rm WD}$. In our calculation, $t \sim
2\times 10^7$ yrs for $M_{\rm WD}=0.25M_\odot$, $\sim 4\times
10^8$ yrs for $M_{\rm WD}=0.2M_\odot$ and up to $\sim 4\times
10^9$ yrs when $M_{\rm WD}=0.15M_\odot$. This explains why so many
very low-mass proto-He WDs have been discovered. For donors with
$M_{\rm 10} \geq 1.60M_\odot$, the $t-M_{\rm WD}$ relation has a
large scatter because the assumption of $M_{\rm c} \simeq M_{\rm
WD}$ is invalid and the value of $M_{\rm env}$ is scattered as
well (see the lower panel of Fig. 8).

\begin{figure}
\includegraphics[width=6.0cm,angle=270]{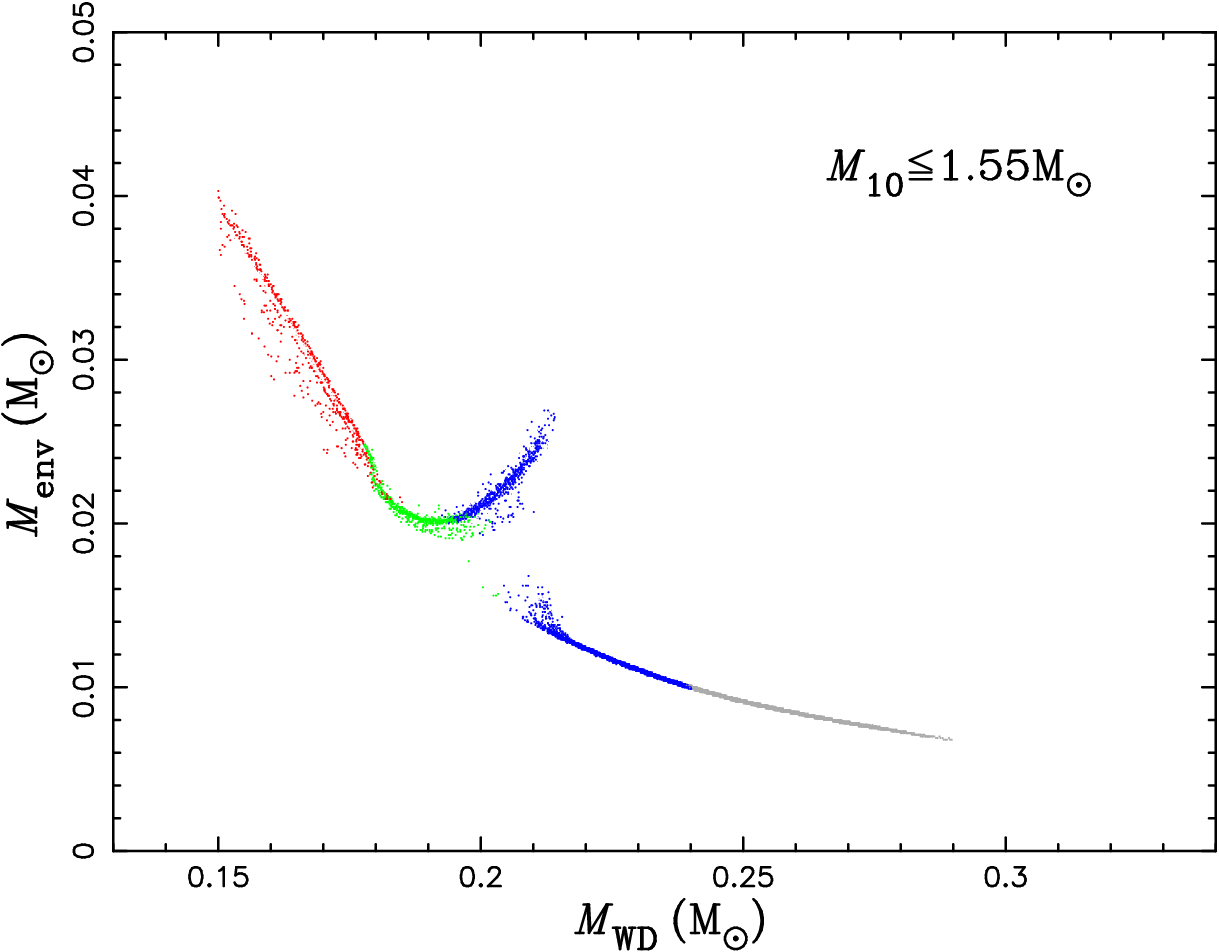}
\includegraphics[width=6.0cm,angle=270]{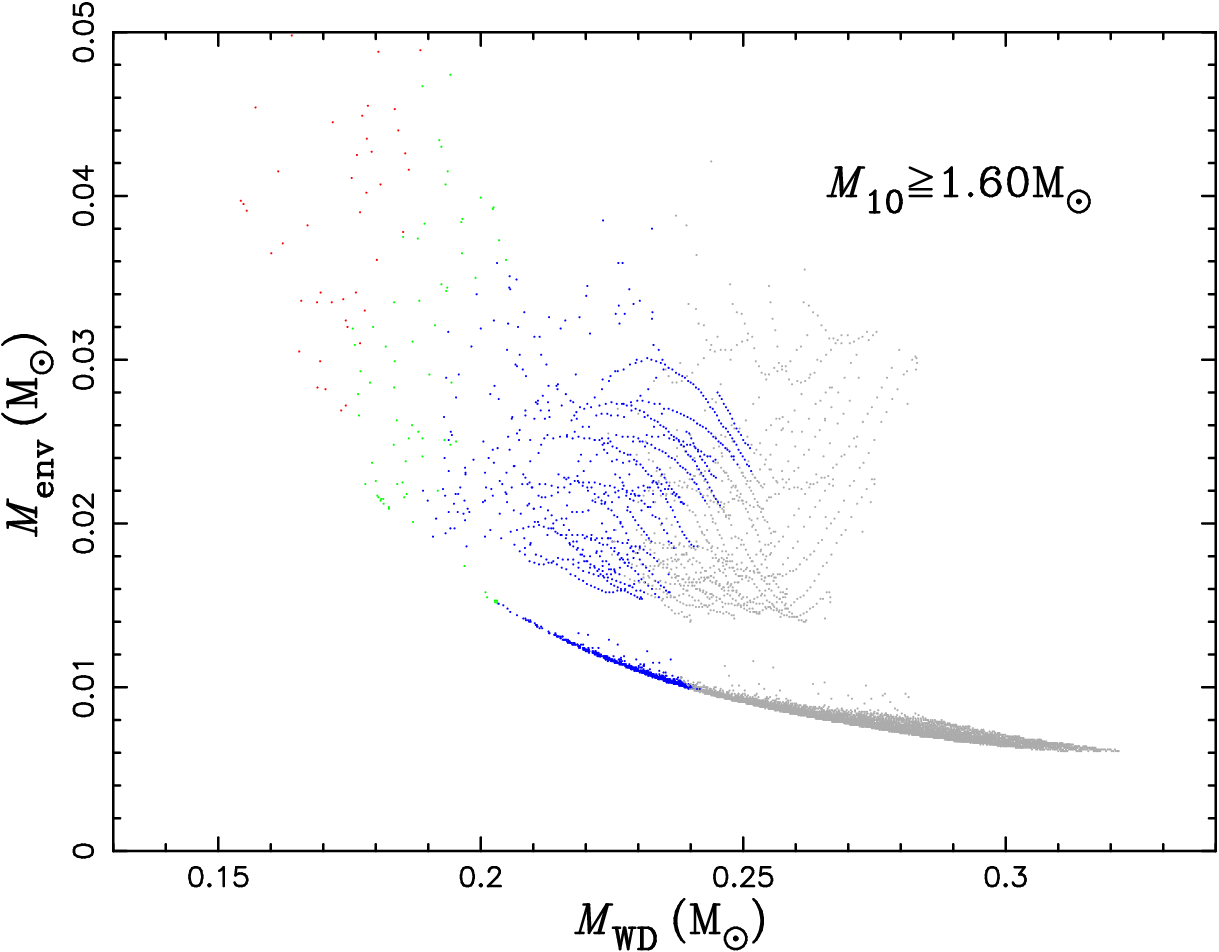}
\caption{The dependance of envelope mass on the proto-He WD mass,
$M_{\rm WD}$, at the end of mass transfer. The colors of red,
green, blue and gray are for the final orbital period in the range
of 0.5--1.0 d, 1.0--2.2 d, 2.2--10 d and $>10$ d, respectively.
The upper panel is for $M_{\rm 10}=0.90-1.55M_\odot$ and the lower
panel for $M_{\rm 10}=1.60-2.00M_\odot$. \label{mfme}}
\end{figure}

The anti-correlation between $M_{\rm env}$ and $M_{\rm WD}$ for
remnants from donors with degenerate cores comes from the
competition between radiation pressure, $P_{\rm r}$, and gravity
pressure, $P_{\rm grav}$, near the end of mass transfer process. A
large $M_{\rm WD}$ indicates a larger core mass $M_{\rm c}$, then
a larger $P_{\rm grav}$ and a higher $T_{\rm b}$ as mentioned
above. So, more shell-burning energy is produced and needs to be
transferred outwards, resulting in a higher $P_{\rm r}$ as well.
When $M_{\rm env}$ is small enough, i.e. the shell-burning energy
can be effectively transferred to the surface and radiate away,
$P_{\rm r}$ reduces dramatically and the giant suffers a sudden
collapse and terminates the mass transfer process. The radiation
pressure dominates the process and leads to the fact that a
massive remnant has a smaller envelope mass. However, the products
appear as two groups as shown in Fig. 8, and there is a small tail
in the upper group, i.e. $M_{\rm env}$ increases with $M_{\rm WD}$
as $M_{\rm WD}\gtrsim 0.19M_\odot$. This is related to the
discontinuous composition gradient produced by the first dredge up
in stars evolving on the RGB. The donor's H-burning shell does not
cross the composition jump during the mass transfer process for
the products on the upper group, but does for those in the lower
group. The tail of the upper group is due to the fact that the
contraction of the donor now is induced by the H-burning shell
crossing the composition jump, rather than $M_{\rm env}$ being
less than a certain value.

For binaries with donors having non-degenerate cores (the lower
panel of Fig. 8), the termination of mass transfer is a gradual
process, i.e. the donor gradually draws back into its Roche lobe.
The envelope mass of the remnant therefore depends on the details
of mass transfer process as well as the initial parameters of each
binary, and results in a big scatter seen in the lower panel of
Fig. 8. In general, for a given $M_{\rm WD}$, a large initial mass
ratio $q_{\rm 0}$ indicates a relatively larger Roche lobe radius
of the donor, then a larger radius and $M_{\rm env}$ for the
remnant at the end of mass transfer. However, we still see a clear
lower $M_{\rm env}$ limit in the figure, which is very similar to
the $M_{\rm env}-M_{\rm WD}$ relation for degenerate donors. The
donors located on the lower limit are generally more evolved than
those beyond the limit, and their cores are degenerate.

\subsection{Comparison with Observations}
To examine the reliability of our calculations for the stable mass
transfer channel for the formation of EL CVn-type stars, we have
compared the results to the observed properties of the current
sample of EL CVn binaries.

\subsubsection{Evolutionary Stage}
Figure 9 shows the radius of the donor in comparison to its Roche
lobe radius at the end of mass transfer, i.e. the onset of the
nearly constant-$L$ phase. Both of the radii are divided by the
orbital separation $a$ for easy comparison with observations,
which are denoted with filled circles. The observations are from
\citet{maxted14a} and the values of $R_{\rm L1}/a$ and the
corresponding error bars are derived from the mass ratios and
their error bars. We see that all of the donor's radii are located
in the range of $0.75-1.0 R_{\rm L1}$ at the end of mass transfer,
and that the donor is closer to $R_{\rm L1}$ for a short orbital
period (see the red points). The ratio of $R_{\rm 1}/R_{\rm L1}$
at the termination of mass transfer indicates a relative density
of the envelope at that time and, for a shorter $P$, the higher
value of $R_{\rm 1}/R_{\rm L1}$ is consistent with a lower $M_{\rm
WD}$ and a larger $M_{\rm env}$ as described in Sect. 3.2. During
the subsequent evolution the value of $R_{\rm L1}/a$ is constant
for each binary, since the separation and the component masses
hardly change during the whole constant-$L$ phase. The models will
then move horizontally to the left in this figure and cross the
observed region for EL CVn-type stars due to the contraction of
$R_{\rm 1}$. 

The onset and end positions of the products for the
nearly constant-$L$ phase on the ($R_{\rm 1}/a, T_{\rm eff}$)
plane are presented in Fig. 10, where several typical evolutionary
tracks are also shown. The theoretical tracks are in general
agreement with the observations. 
When $M_{\rm 10}\ge1.60M_\odot$, 
the theoretical evolutionary tracks may well match the location of the EL CVn samples, but
the parameter space for short orbital-period EL CVn stars ($P\le2.2$ d) is very small (see section 3.1). 
It is unlikely that most of the observed EL CVn stars are produced from the systems with $M_{\rm 10}\ge 1.60M_\odot$. 
For the models of $M_{\rm 10}\le1.55M_\odot$, the observed systems tend to be
~1000 K hotter than the predicted $T_{\rm eff}$ values. For most
of these systems the $T_{\rm eff}$ estimates are based only
photometric data and the proto-He WD contributes only 5-10 per cent
of the light from the binary at most wavelengths, so it may be
that this offset is due to a bias in the method used to estimate
$T_{\rm eff}$ of these stars. The sample of EL CVn-type stars were
identified by looking for short-period light curves with a
characteristic `boxy' shape (sharp ingress and egress) that is
deeper than the more rounded secondary eclipse. This indicates
that the smaller star in the binary is hotter and so must be a
highly evolved star. Near the onset of the constant-$L$ phase, the
proto-He WD has a similar radius and effective temperature to its
companion. The light curve of such a binary shows strong
ellipsoidal effect and smoother eclipse, quite similar to the
light curve of W UMa-type contact binaries. Indeed, AW UMa is an
example of a binary system with a W UMa type light curve that has
subsequently been found to be a binary containing a proto-He WD
(Pribulla \& Rucinski 2008, but see Eaton 2016 for an alternative
interpretation of this system).
The Kepler object KIC 8262223 is a good candidate
 being in the onset of the constant-$L$ phase according to its orbital parameters. We see obviously the ellipsoidal effect in its light curve but without the `boxy' shape \citep{guo16}. Near the end of the constant-$L$
phase the eclipses become shallow and difficult to spot in
ground-based light curves, but several systems have been
identified using photometry from the Kepler space craft
\citep{rap15,carter11,breton12,fai15}.

\begin{figure}
\includegraphics[width=6.0cm,angle=270]{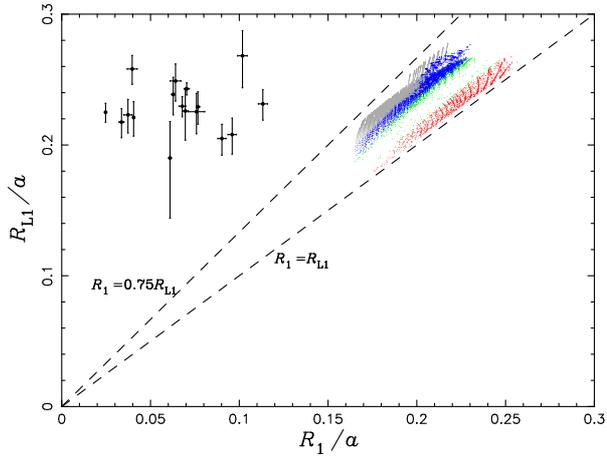}
\caption{The donor's radius ($R_{\rm 1}$) versus its Roche lobe
radius ($R_{\rm L1}$) at the end of mass transfer. Both of the
radii are divided by the orbital separation $a$ for easily
comparing with observations. The colors of red, green, blue and
gray are for the final orbital period in the range of 0.5--1.0 d,
1.0--2.2 d, 2.2--10 d and $>10$ d, respectively. The dashed lines
are for $R_{\rm 1}=R_{\rm L1}$ and $R_{\rm 1}=0.75R_{\rm L1}$ as
indicated in the figure. The filled circles are observed EL
CVn-type stars from \citet{maxted14a} where the values of $R_{\rm
L1}/a$ and the corresponding error bars are derived from the mass
ratios and their error bars. In following evolution, the models
will move straightforwardly to the left and cross the observed
region for EL CVn-type stars due to the contraction of $R_{\rm
1}$. \label{rlrg}}
\end{figure}

\begin{figure}
\includegraphics[width=6.0cm,angle=270]{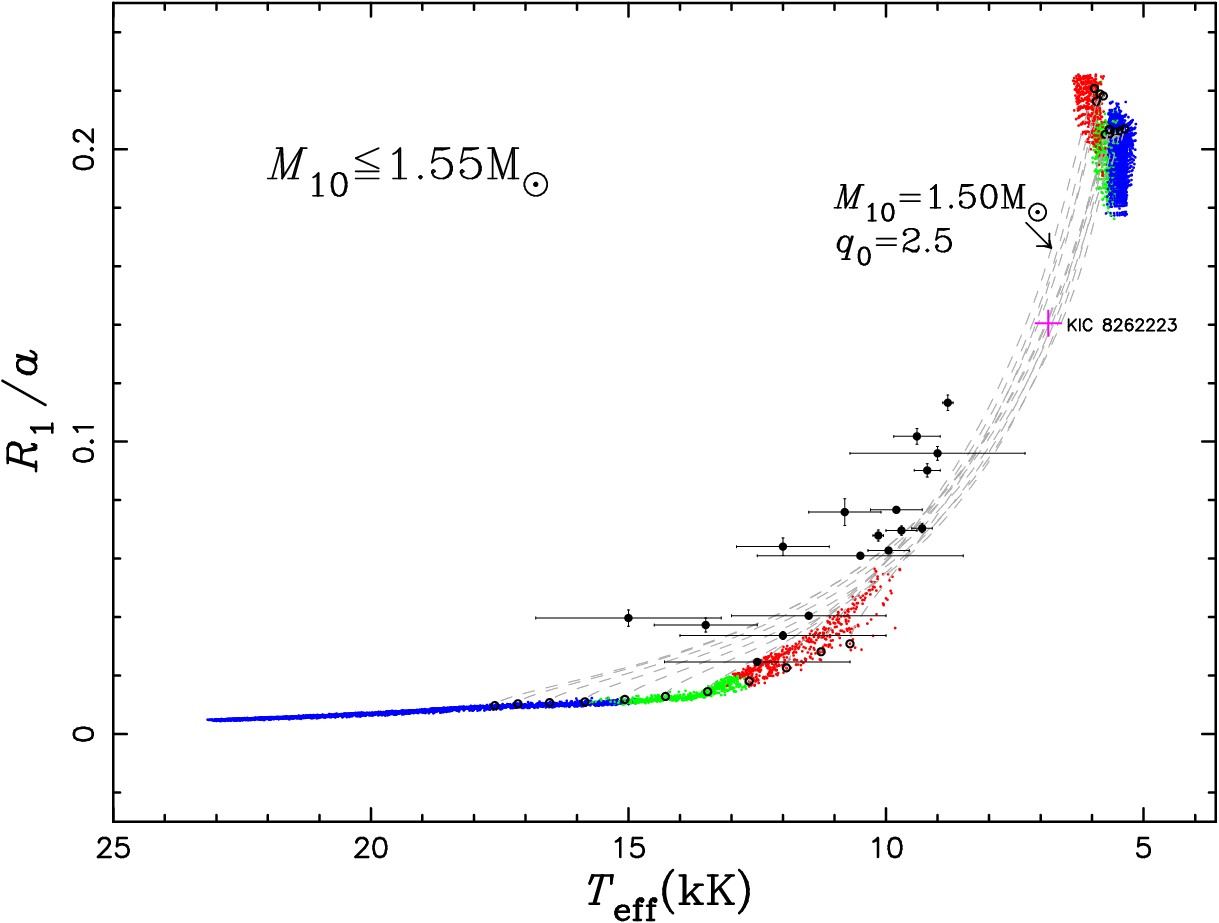}
\includegraphics[width=6.0cm,angle=270]{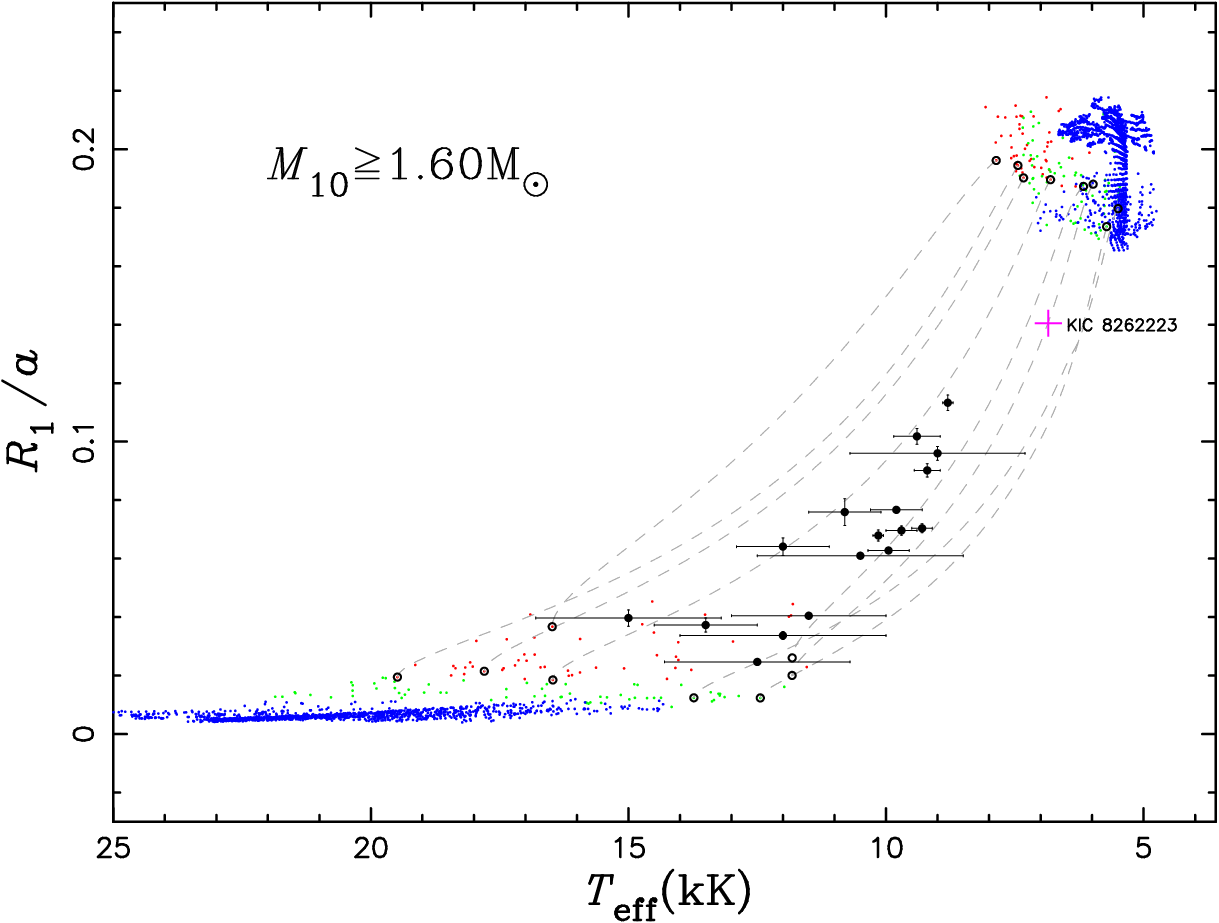}
\caption{Evolutionary tracks of the proto-He WDs in the ($R_{\rm
1}/a, T_{\rm eff}$) plane, where $a$ is the orbital separation,
$R_{\rm 1}$ and $T_{\rm eff}$ are the radius and the effective
temperature of the proto-He WDs, respectively. For clarity, we only
present the onset (dots, the upper right group) and end points
(dots, the lower left group) for the nearly constant-$L$ phase and
several typical evolutionary tracks during this phase (the
starting and ending places are indicated with open circles).
Various colors are for different orbital period ranges, i.e. the
red, green, blue and gray are for the final orbital period in the
range of 0.5--1.0 d, 1.0--2.2 d, 2.2--10 d and $>10$ d,
respectively. The filled circles are the observed EL CVn-type
stars from \citet{maxted14a}. 
The pink plus indicates the location of KIC 8262223 from \citet{guo16}.  \label{track}}
\end{figure}

\subsubsection{The Mass -- Period Relation}
As introduced in Sect. 1, a fairly tight relation exists between
the final orbital period and the WD mass if the WD evolves from
stable mass transfer and its progenitor has a degenerate core.
Here we show this relation in the upper panel of Fig. 11, where the
black dots are from our models and the solid line is for the
fitting formula of \citet{lin11}, that is,
\begin{equation}
P\simeq \frac{4.6\times10^6M_{\rm WD}^9}{(1+25M_{\rm
WD}^{3.5}+29M_{\rm WD}^6)^{3/2}} {\rm (d)}.
\end{equation}
Our result is quite similar to that of \citet{lin11}. There is an
obvious narrow band on which the descendants of donors with
degenerate cores have gathered. The systems deviating from the
band originate from donors with non-degenerate cores which do not
follow the core mass-radius relation for giants.
The Kepler objects and EL CVn stars with known orbital periods and
WD masses are marked in the figure. The agreement between the
observations and the results from our grid of models is
satisfactory given that there has been no attempt to optimize the
initial parameters of the binary (e.g., metallicity) or the
details of the evolution to match the observed properties of the
individual binary systems.

We notice a small bulge for $M_{\rm WD}\lesssim0.22M_\odot$ on the
$M_{\rm WD}-P$ band, which results from relatively large stellar
radii due to the fact that the cores ($M_{\rm
c}\lesssim0.19M_\odot$) in such low-mass stars are non-degenerate
at the termination of mass transfer. 
For $Z=0.004$ (see the lower panel of Fig.11), the bulge appears when $M_{\rm
WD}\lesssim0.25M_\odot$ (or $M_{\rm c}\lesssim0.22M_\odot$) due to 
the higher He abundance in the core at a low $Z$.
We see a similar bulge (appeared at the proto-He WD mass of $0.21M_\odot$) 
in Fig.13 of \citet{ist14a}.
Accompanying the bulges, the $M_{\rm WD}-P$
relation has a certain degree of dispersion, consistent with
non-degenerate cores of the donors in this mass range. We also
notice that the appearance of the bulges reduces the difference of
the relation induced by the metallicity: products with low-mass He
cores from various metallities likely mix together and become
undistinguishable, different from those with more massive He
cores.

As mentioned in section 1 and studied by many authors in literature, 
the $M_{\rm WD}-P$ relation is determined by the core mass - luminosity relation of stars 
with degenerate cores, and it is hardly affected by the assumptions in the modeling of binary evolution.
In the low mass end, however, the cores are not degenerate yet at the termination of mass transfer. 
We then checked the influence of assumptions adopted in binary evolution on the $M_{\rm WD}-P$ relation.
The results are shown in Fig.12, in which we include the products in the following cases (1) models from our model grid with $M_{\rm 10}\le 1.55M_\odot$ (the black dots), (2) mass transfer is completely conservative (no mass and angular momentum loss, the red dots), (3) mass transfer is completely non-conservative and the lost mass takes away the specific angular momentum of the donor (the grey crosses), (4) 50 per cent of the lost mass from the primary leaves the system but does not take away any angular momentum (the filled green circles), and (5) similar to case (4) but all the mass lost from the primary leaves the system (the filled red circles)\footnote{The former three cases are usually adopted in the modeling of binary evolution for different objects e.g \citet{nel01,han02}, while the last two are not common in the study of binary evolution. We include the last two cases here to see the maximum effect in some extreme assumptions.}.

From Fig.12 we see that the products from the former three cases are well overlapped, and those from the last two are very close to them, indicating that the assumptions in the modeling of binary evolution (even in some extreme assumptions) have very little influence on the  $M_{\rm WD}-P$ relation. In fact, if we assume no angular momentum being taken away by the lost mass, as for the last two cases here, the change of the orbital separation becomes small and the termination of mass transfer is more dependent on the structure of the donor than before, and the resulting binary may agree with the $M_{\rm WD}-P$ relation better. 

Moreover, when we plot all the products on Fig.6 and Fig.8, we find that they are undistinguishable with those of our model grid calculations, indicating that they have similar lifetimes and envelope masses at a certain proto-He WD mass. It is easily understood by the fact that, from the theory of stellar structure and evolution, whether a star contracts (mass transfer terminates) or not is only determined by its intrinsic structure (e.g. chemical composition), independent of detailed mass loss process. From this point of view, we may expect a similar minimum proto-He WD mass from various assumptions in binary evolution at a given chemical composition. 

Based on the $M_{\rm WD}-P$ relation, we estimate that the first
two `bloated hot WDs' discovered using Kepler photometry have
masses in the range $0.26-0.27M_\odot$ for KOI 81b and
$0.205-0.25M_\odot$ for KOI 74b. The relatively large span of the
mass range for KOI 74b is due to the fact that the donors with
non-degenerate cores also leave remnants with such an orbital
period ($P=5.18875$ d) but various masses at the termination of
mass transfer (see the discussion in Sect. 3.2). It is around
$0.22M_\odot$ if we only consider the contribution from donors
with degenerate cores. Both of the masses for the two objects
match the result of \citet{van10}, where the mass of KOI 74b is
derived from Doppler boosting and the $M_{\rm WD}-P$ relation of
\citet{rap95}. With the constraint on orbital period in
\citet{maxted14a}, i.e. $P\le2.2$ d, all the discovered EL CVn
stars have masses $\le \sim 0.2M_\odot$, so remain discoverable
for much longer lifetimes than more massive proto-He WDs.

\begin{figure}
\includegraphics[width=6.0cm,angle=270]{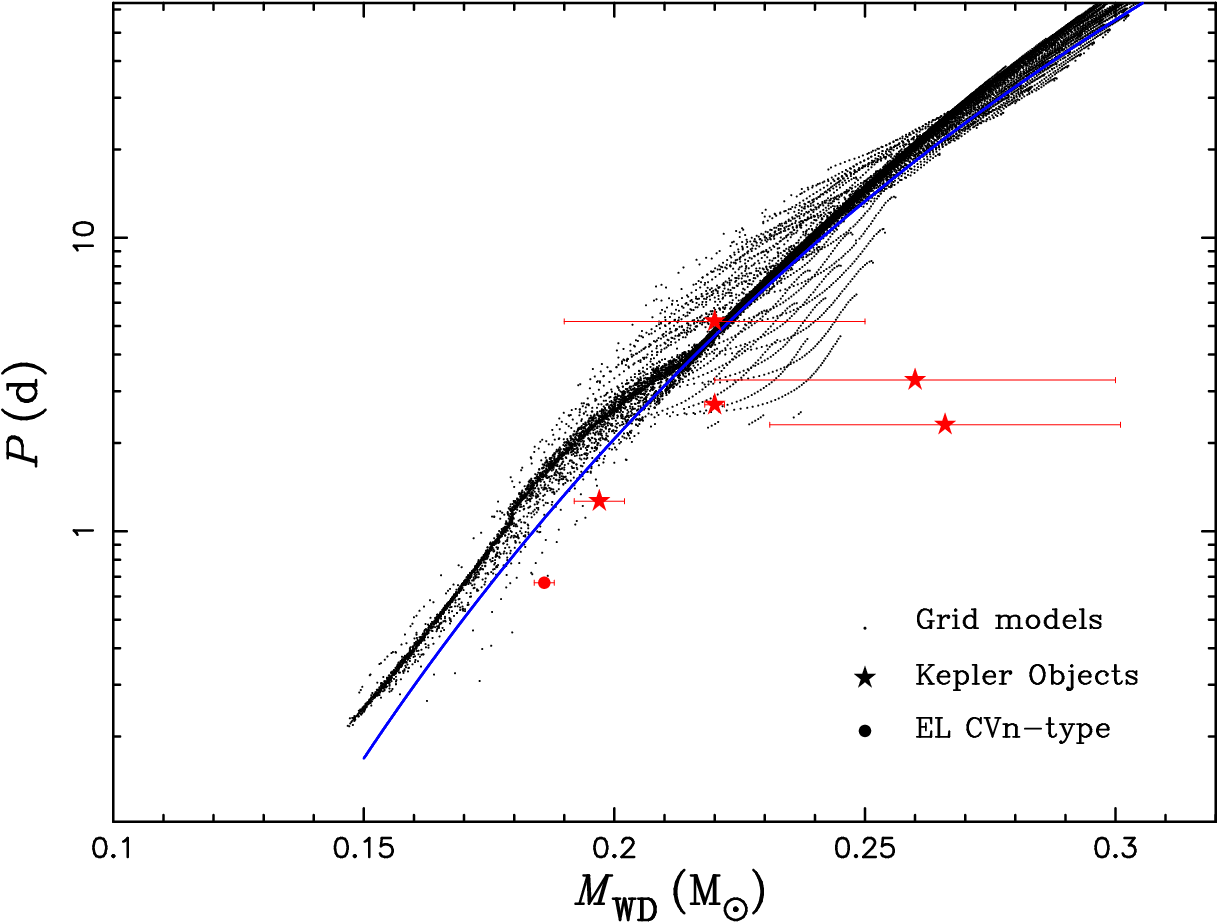}
\includegraphics[width=6.0cm,angle=270]{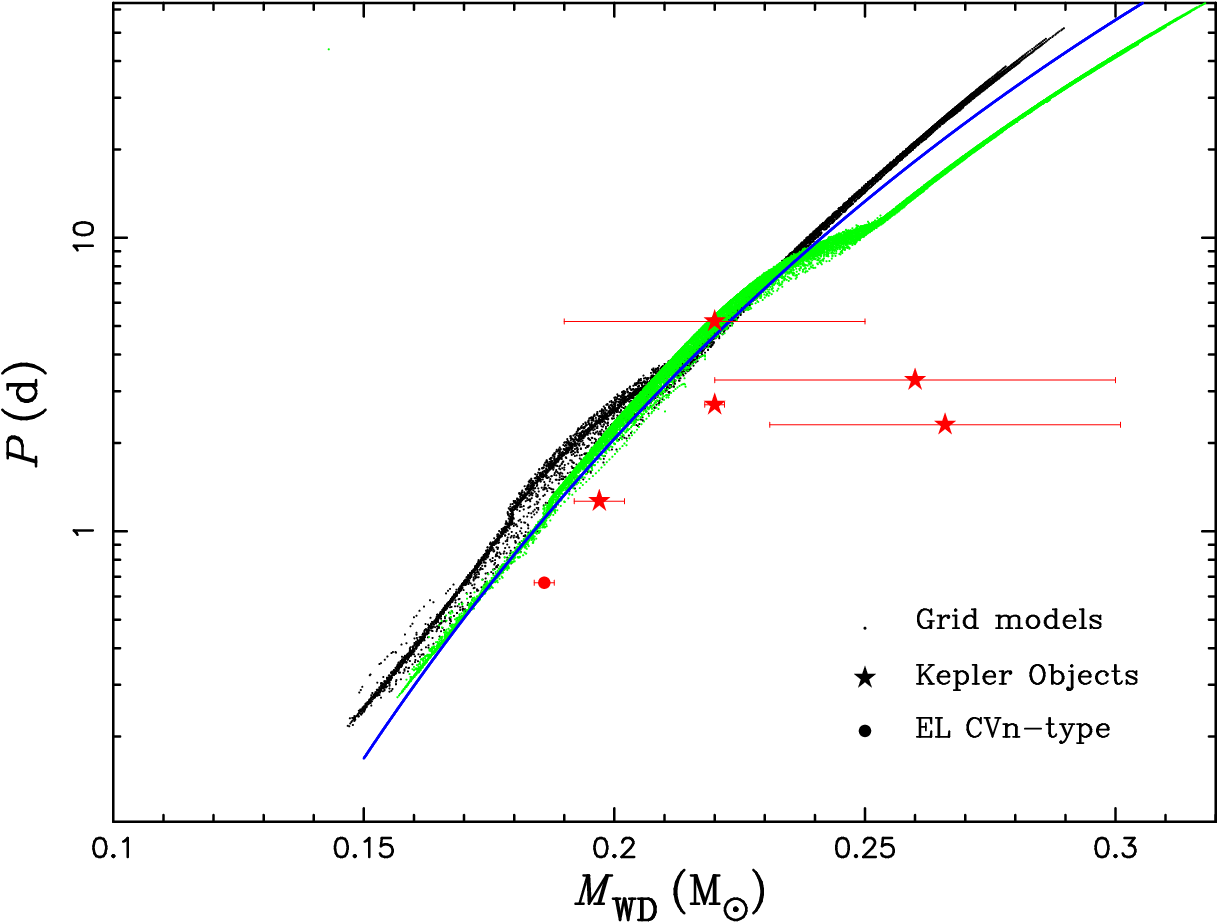}
\caption{The orbital period--WD mass relation obtained from stable
mass transfer. The upper panel shows all the products (black dots)
in our calculations. The filled circles and stars are for EL
CVn-type stars \citep{maxted13,maxted14b} and Kepler objects
\citep{van10,carter11,breton12,rap15}, respectively. The fitting
formula from \citet{lin11} is also presented for comparison. The
lower panel is only for $M_{\rm 10}\le 1.55M_\odot$. The black
dots are for $Z=0.02$ and the green ones are for $Z=0.004$ for
comparison. \label{mfpf}}
\end{figure}

\begin{figure}
\includegraphics[width=6.0cm,angle=270]{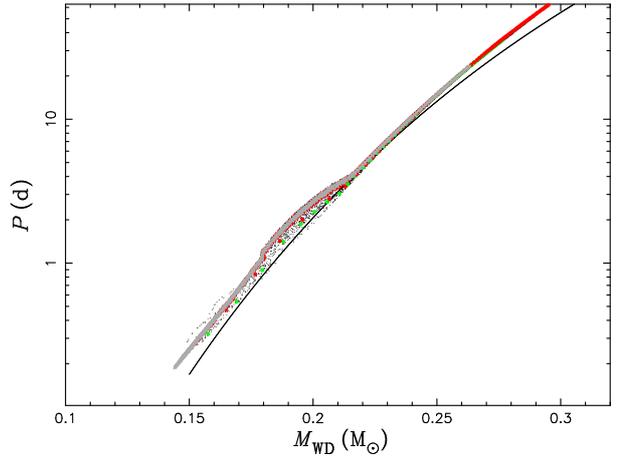}
\caption{The orbital period-He WD mass relation from various assumptions in the 
modeling binary evolution. The black dots are for the models of our grids with 
 $M_{\rm 10}\le 1.55M_\odot$, the red dots and grey crosses, overlapped on the black dots, are 
 for those from completely conservation and completely non-conservation cases, respectively, 
 and the green and red filled circles are for those 
 from the assumptions that  50 per cent and 100 per cent of the lost mass from the primary 
 leaves the system but does not take away any angular momentum, respectively.   \label{p-mass}}
\end{figure}

\section {Binary Population Synthesis}
We perform a Monte-Carlo simulation to obtain a stellar
population, then interpolate in the grid above to obtain binary
proto-HeWDs produced from stable mass transfer. In the Monte Carlo
simulation, all stars are assumed to be members of binaries and
have circular orbits. The primaries follow the initial mass
function of \citet{miller79} and are generated according to the
formula of \citet{egg89}, in the mass range 0.08 to $100M_\odot$.
The secondary mass, also with a lower limit of $0.08M_\odot$, is
obtained from a constant mass-ratio distribution $n(q)=1$. Two
other mass ratio distributions, $n(q)=2q$ and uncorrelated
component masses (i.e., the component mass is given in a way the
same as that of the primary), are examined for estimating the
uncertainty induced by the mass-ratio distribution. The
distribution of orbital separations $a$ is taken to be constant in
${\rm log }$ {\it a} for wide binaries. This separation
distribution has been used in many Monte-Carlo simulations and
implies an equal number of wide binary systems per logarithmic
interval and approximately 50 per cent of stellar systems with
orbital periods less than 100 yr \citep{chen09}.
Long-orbital-period binaries are effectively single stars. To
mimic the case of the Galaxy, we assume a constant star formation
rate of $5M_\odot {\rm yr}^{-1}$ \citep{yung98,wb13}. Table 1
gives a brief summary of the parameters.

\begin{table}
\caption{Initial parameters for binary population synthesis. IMF--
initial mass function, $q$--mass ratio (the primary/the
secondary), $a$--separation. }
\begin{tabular}{ll}
\hline
IMF for the primary & Miller\& Scalo (1979) \\
& $n(q)=1$ (standard)\\
Mass-ratio distribution &$n(q)=2q$ \\
& uncorrelated\\
Separation distribution & $n({\rm log}a)$=const \\
Star formation rate & $5M_\odot {\rm yr}^{-1}$\\
\hline \label{bps}
\end{tabular}
\end{table}

\subsection{Distributions of Binary Parameters}
\begin{figure}
\includegraphics[width=6.0cm,angle=270]{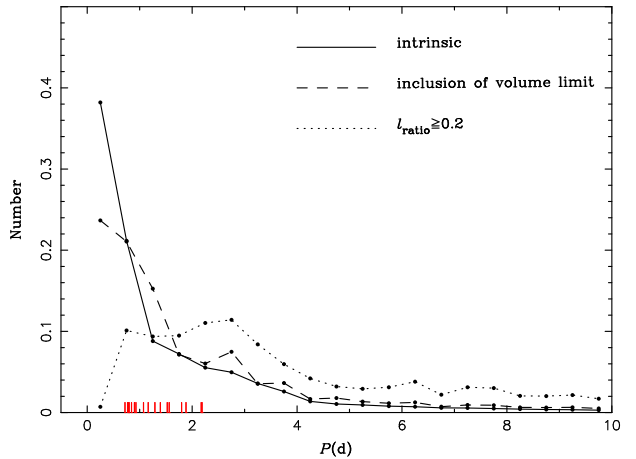}
\caption{Orbital period distributions of EL CVn-type stars from
the simulation of $n(q)=1$ (the standard case, normalized). The
solid line is intrinsic while selection effects are simply
included for the dashed and dotted lines, i.e. the dashed line is
for the inclusion of volume limit simply based on the luminosity
of A- or F-type component, and the dotted line is for the
constraint on the luminosity ratio of both components superimposed
on the volume limit (see text for the details). Short ticks along
the $x$-axis indicate the positions of EL CVn stars in Table 2 of
\citet{maxted14a}. \label{p-dis}}
\end{figure}

\begin{figure}
\includegraphics[width=6.0cm,angle=270]{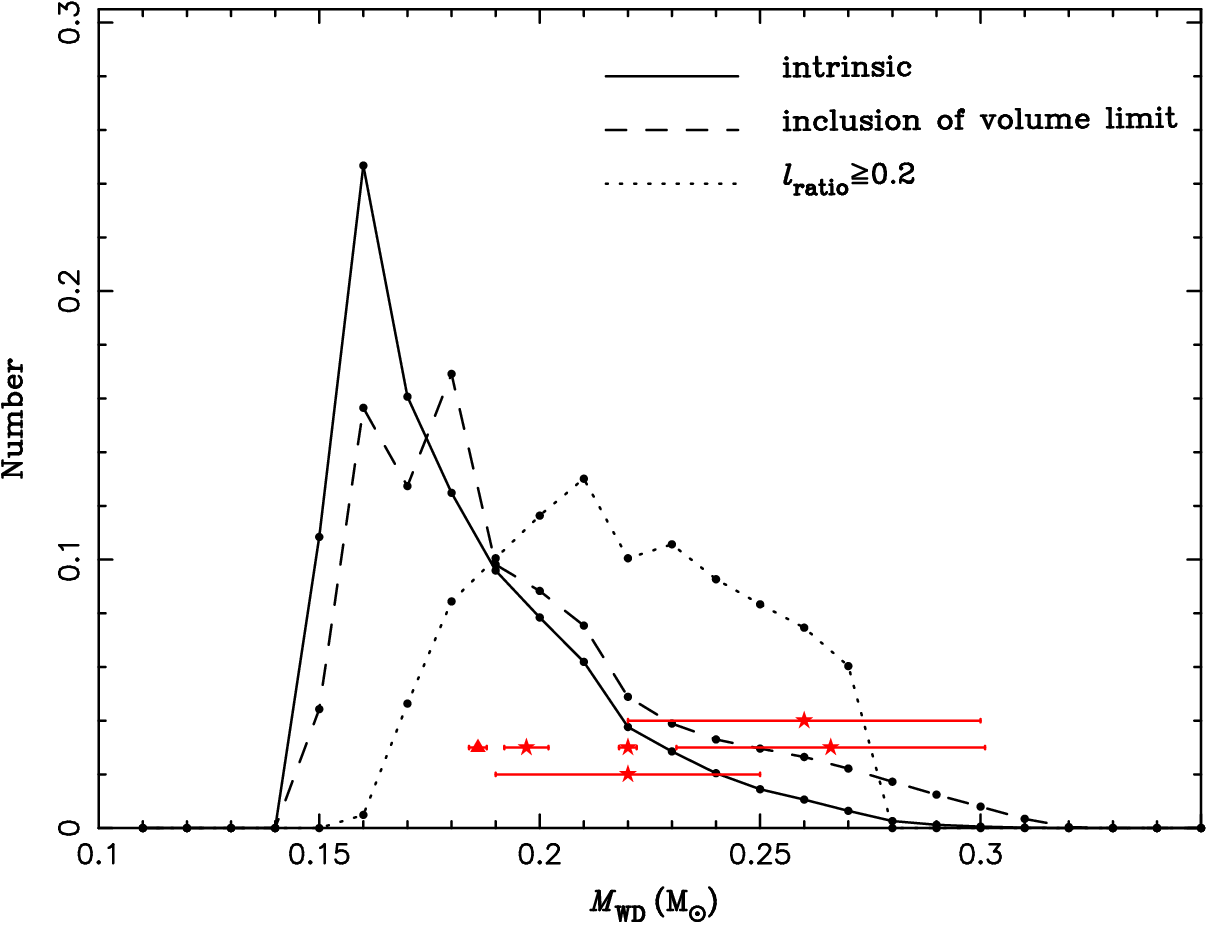}
\caption{Similar to Fig.13 but for the WD mass distributions. The
filled triangles and stars are for EL CVn-type stars
\citep{maxted13} and Kepler objects
\citep{van10,carter11,breton12,rap15}, respectively.
\label{mhe-dis}}
\end{figure}

\begin{figure}
\includegraphics[width=6.0cm,angle=270]{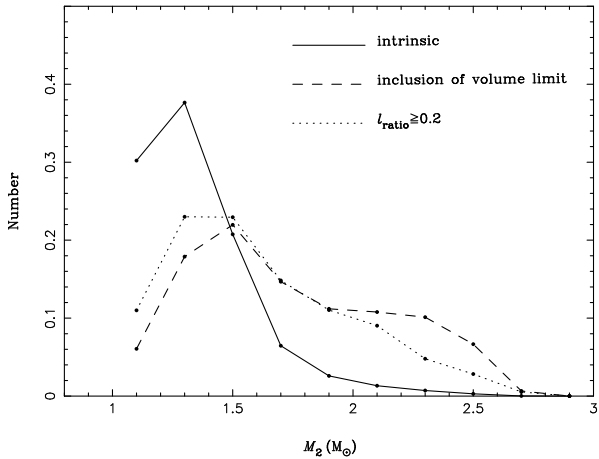}
\caption{Similar to Fig.13 but for the companion mass.
\label{m2-d}}
\end{figure}

Figures 13--15 show the distributions of orbital period $P$, the
WD mass $M_{\rm WD}$ and the companion mass $M_{\rm 2}$,
respectively, for EL CVn-type stars at 13.8 Gyr from the
simulation of $n(q)=1$ (the standard model). Different initial
mass ratio distributions give similar results on the three binary
parameters. In each of the figures, the black solid line is
intrinsic while some selection effects are included for the dashed
and dotted lines. For the dashed lines, we simply multiply a
factor of $L_{\rm A}^{3/2}$ to include the volume limit induced by
the luminosity. Here $L_{\rm A}$ is the luminosity of the A- or
F-type dwarf companion and is given by the classical
mass-luminosity relation of MS stars (Eq(3)). The most typical feature for
EL CVn-type stars is the boxy eclipse of the light curve, so the
most important selection effect probably comes from the factor
that prevents us to discover this feature, e.g. the luminosity
ratio of both components in the WASP band (400-700nm). The boxy
eclipse becomes very shallow and difficult to find if the
luminosity of proto-He WD in this band is far below its companion's.
However, it is a little bit difficult to obtain the luminosity of
the proto-He WD components in this band since they pass through a
very large effective temperature range. We therefore artificially
choose the luminosity of the proto-He WDs at $T_{\rm eff}=10000$ K,
$L_{\rm He}$, compare it with $L_{\rm A}$, and assume that the
objects with $L_{\rm He}/L_{\rm A}<0.2$ (we referred the case of
WASP J0247-25) cannot be discovered since the depth of the eclipse
is shallow and the boxy property is not obvious on the light
curves. This effect is superimposed to the luminosity limit of the
companion and the result is shown as dotted lines in the figures.
Though the considerations of the selection effects here are very
simple, but they still give us some useful information\footnote{In Fig.14
we have included the Kepler systems though they suffer from different selection effects than those analyzed in the figure since we have only one EL CVn star which has the exact proto-He WD mass presently. }. 

For the intrinsic distributions, since low-mass proto-He WDs have
much longer lifetimes and shorter $P$ as discussed in Sect. 3,
most EL CVn-type binaries have proto-He WDs with mass around the
minimum mass allowed for such objects, i.e. the proto-He WDs have a
mass peak around $0.16M_\odot$ (Fig. 14) and the corresponding
orbital period peak is 0.2 d (Fig. 13). These distributions are
unlikely to be affected by the treatment of mass transfer process
since whether a star evolves across the constant-$L$ phase or ever
decreasing mass is only related to its intrinsic structure, which
is determined by initial chemical composition and descriptions of
the basic theory of stellar structure and evolution. The
consideration of volume limit shifts the WD mass peak to $\sim
0.18M_\odot$, and the constraint on the luminosity of both
components further shifts the peak to $0.21M_\odot$. The orbital
period distribution changes accordingly. We see that, with the
inclusion of both of the selection effects, the WD mass
distribution becomes more flat, indicating similar possibilities
in a relatively wide WD mass range. Fig. 13 shows that the observed
EL CVn-type binaries are well located in the orbital period range
given by the two selection effects (the dotted line). The
companion mass peak also changes with selection effects, however,
this distribution is more likely to be affected by the assumption
of mass loss during mass transfer (see discussion in Sect. 5).

We can see more details from 2D distributions of the binary
parameters, i.e. the ($P,q$) plane and the ($M_{\rm WD}, M_{\rm
2}$) plane, as shown in Figs. 16--17. There are four panels in
both figures, where panel (a) is for the case at birth and panel
(b) for the inclusion of weight of lifetime. In panel (c) we
simply multiply a factor of $L_{\rm A}^{3/2}$ and in panel (d) we
further assume the objects with $L_{\rm He}/L_{\rm A}<0.2$ cannot
be discovered based on panel (c). We firstly see the ($P,q$)
plane, which can be directly compared with observations. Note that
most of the mass ratio estimates in this figure are based on the
amplitude of the ellipsoidal effect in the light curve. These will
be biased towards higher values of $q$ if this amplitude is
under-estimated, e.g., due to contamination of the light curve by
light from a third star in the system or systematic errors in the
photometry. The theoretical EL CVn-type stars are distributed over
a wide range of period at birth (panel (a)), but gather at the
short-$P$ end due to long lifetimes of low-mass proto-He WDs (panel
(b)), similar to the solid line in Fig. 13. The consideration of
volume limit from the companion's luminosity has not improved the
situation significantly (panel (c)). However, if we further
include the constraint on the luminosity ratio of both components,
the observation may be well reproduced (panel (d)) except for
those with very shorter $P$, i.e. less than 1 d. Such short-$P$ EL
CVn-type stars are likely from the less-conservative mass transfer
process than that assumed in this paper. The gray scale image of
($M_{\rm WD}, M_{\rm 2}$) shows that the most common component
masses of EL CVn-type stars are $M_{\rm WD}=0.17-0.21M_\odot$ and
$M_{\rm 2}=1.3-1.5M_\odot$, and that the object WASP J0247-25 is
well located in the most common region. This is also consistent
with the observations that the brighter component in most EL
CVn-type binaries is typically an A-type dwarf.

\begin{figure}
\includegraphics[width=6cm,angle=270]{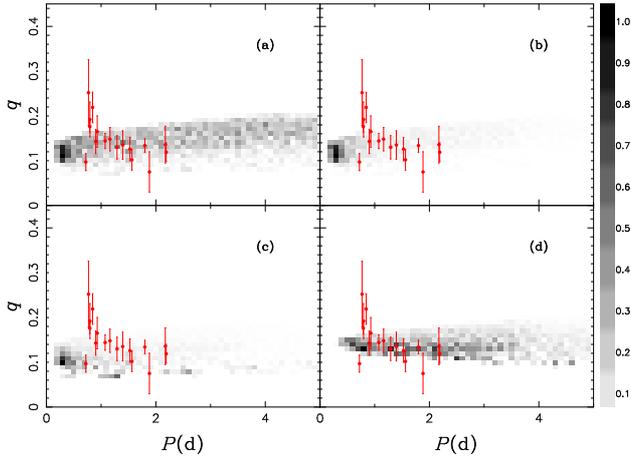}
\caption{Distribution of EL CVn-type stars in the ($P,q$) plane,
where $q$ is mass ratio and $P$ is orbital period. Panel (a) is
for the case at birth and panel (b) for the inclusion of weight of
lifetime in the nearly constant-$L$ phase. Selection effects are
simply included in panels (c) and (d), i.e. panel (c) is for the
volume limit based on the luminosity of A- or F-type component,
and panel (d) further includes the constraint on luminosity ratio
of both components in addition to the volume limit (see text for
the details). \label{pq-dis}}
\end{figure}

\begin{figure}
\includegraphics[width=5.8cm,angle=270]{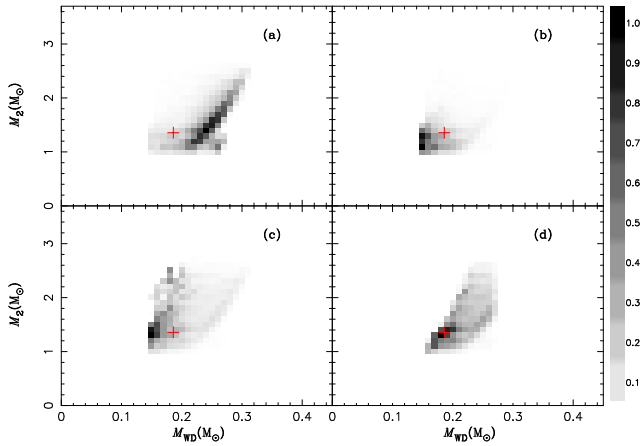}
\caption{Similar to Fig.16 but in the ($M_{\rm WD}, M_{\rm 2}$)
plane, where $M_{\rm WD}$ is the proto-He WD mass and  $M_{\rm 2}$
is the mass of the A- or F-type component. The plus is for WASP
J0247-25 \citep{maxted13}\label{m1m2-dis}}
\end{figure}

\subsection{Birthrates and Numbers}

\begin{figure}
\includegraphics[width=6.0cm,angle=270]{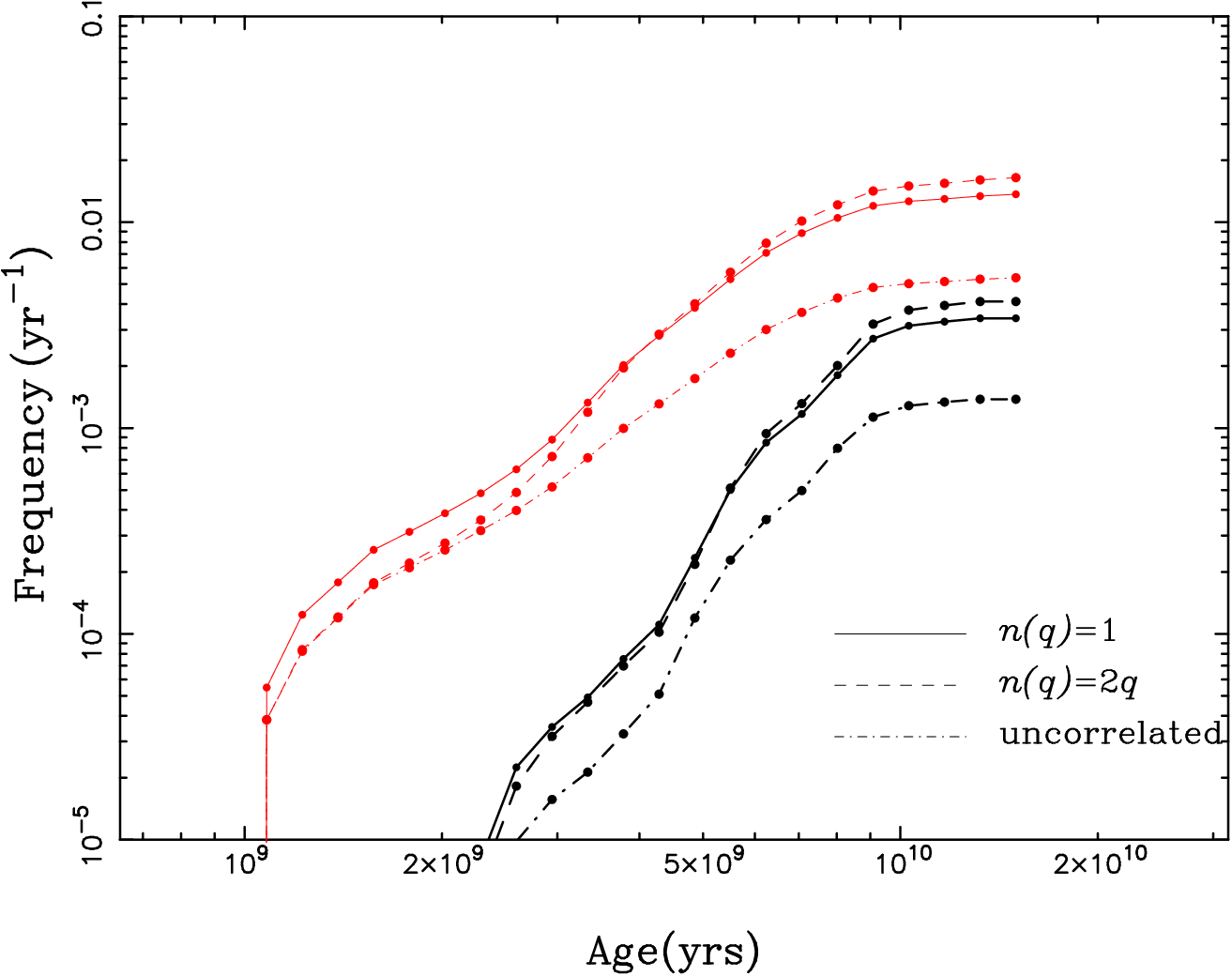}
\includegraphics[width=6.0cm,angle=270]{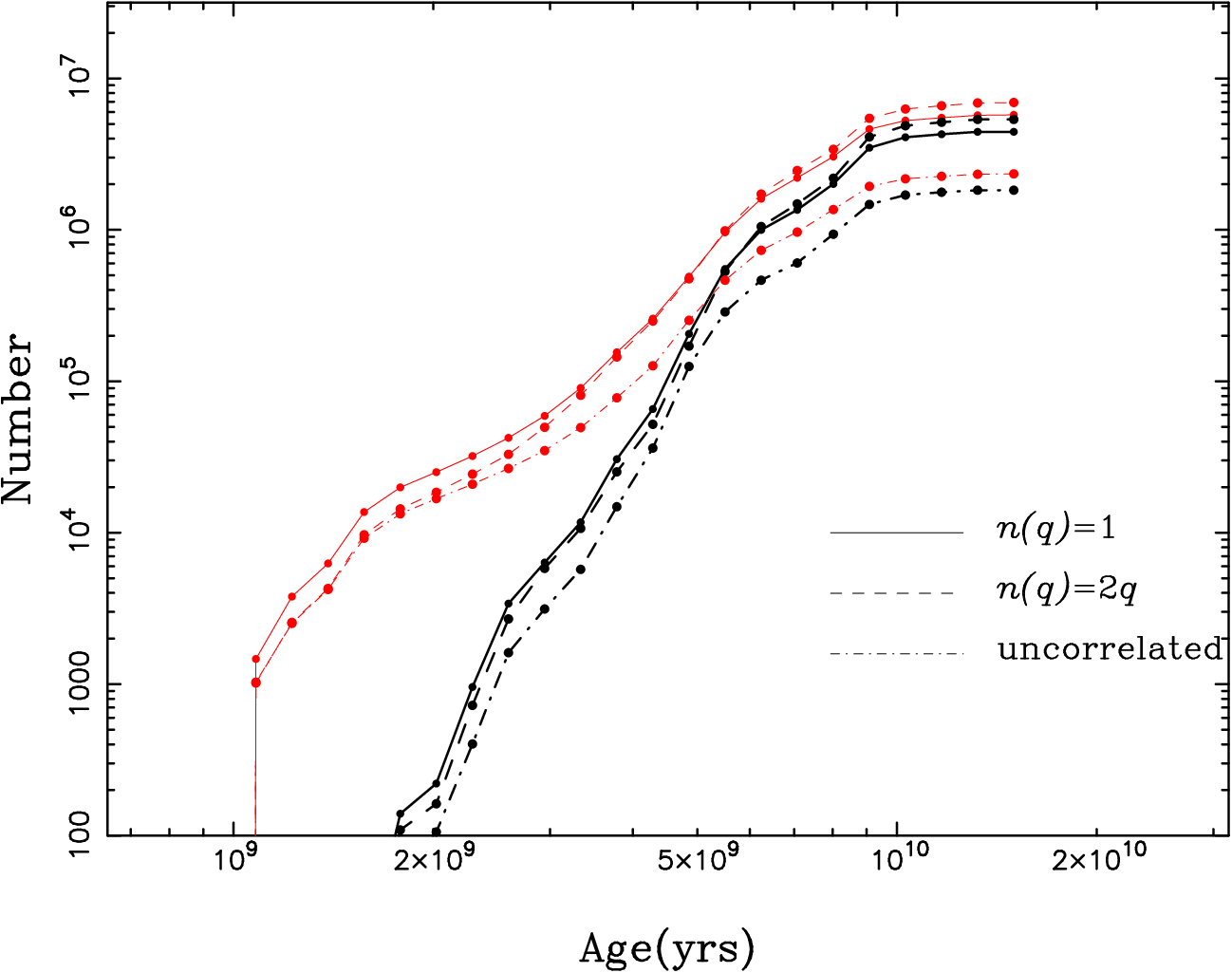}
\caption{Birthrates (the upper panel) and numbers (the lower panel)
of EL CVn-type stars for a constant star formation rate of
$5M_\odot{\rm yr}^{-1}$. The upper three red thin lines are for
all the products with orbital period $P\leq 10$ d, and the bottom
three black thick lines are just for those with $P\leq 2.2$ d.
Different line styles are for different mass ratio distributions
as indicated in the figure (see text for details).
\label{const-birthrate}}
\end{figure}

The birthrates for EL CVn-type stars are shown in the upper panel
of Fig. 18, where the upper three red thin lines are for $P\le 10$
d (referred as $f_{\rm 10}$) and the bottom black thick ones for
$P\le2.2$ d (referred as $f_{\rm 2.2}$). Both $f_{\rm 2.2}$ and
$f_{\rm 10}$ increase with age and are nearly constant for ages
$>\sim 9\times10^9$ yrs, where $f_{\rm 2.2}\simeq0.0015$, $f_{\rm
10}\simeq0.005$ for the uncorrelated mass ratio distribution and
$f_{\rm 2.2}\simeq0.004$, $f_{\rm 10}\simeq0.015$ for the other
two mass ratio distributions. The corresponding numbers of EL
CVn-type stars are shown in the lower panel of the figure. Since
the most common donor mass for producing EL CVn-type stars with
$P\le 2.2$ d are in the range $1.15-1.2M_\odot$ (Figs. 3--5) and
those with short $P$ have much longer lifetimes, short
orbital-period EL CVn-type stars become more and more important in
all of such binaries, e.g. the number of EL CVn-type stars with
$P\le 2.2$ d is $2\times 10^6$ from a total population of
$2.5\times10^6$ such binaries with $P\le 10$ d for the
uncorrelated mass ratio distribution, or $5\times 10^6$ in a total
of $6\times 10^6$ for the other two. The local density is then
$4-10\times 10^{-6}{\rm pc}^{-3}$ for EL CVn-type stars with
$P\le2.2$ d (the volume of the Galaxy is adopted as $5\times
10^{11} {\rm pc}^3$). For populations younger than 2.5 Gyrs,
$f_{\rm 2.2}\leq 10^{-5}{\rm yr}^{-1}$, which results in less than
1000 EL CVn-type stars with $P\le2.2$ d and a local density less
than $2\times 10^{-9} {\rm pc}^{-3}$ under various mass ratio
distributions. Therefore, many more low-mass EL CVn-type stars
remain to be discovered, and these stars will mostly belong to old
stellar populations. Only 8 of the known EL CVn-type binaries
currently have measurements of their Galactic kinematics
\citep{maxted14a}, but 2 of them are members of old stellar
populations (halo or thick disc). The velocity dispersion of the
other 6 stars is $\sigma_v\approx45$\,km\,s$^{-1}$, suggesting a
mean kinematic age for this sub-sample $\approx 4$\, Gyr
\citep{cox00}. This is consistent with our prediction that the
Galactic population of EL CVn-type stars is dominated by old
stars.

\section{Discussions}
\subsection{The assumption of non-conservative mass transfer}
As mentioned in section 2, non-conservative mass transfer is among the
major uncertainties in binary evolution. We have discussed that some of the intrinsic properties
of the proto-He WDs in EL CVn stars, such as the minimum mass, the envelope mass, lifetime, evolutionary track, and the $M_{\rm WD}-P$ relation, will not change with the assumptions in binary evolution in section 3.3.2. However, the parameter space for producing EL CVn-type stars, their predicted birthrate and the
number of such objects do change with what we assume for
the binary evolution. Here we examined two other assumptions 
in the modeling binary evolution, that is, the mass transfer process 
is completely conservative (all the mass lost from the donor is accreted by the companion)
or completely non-conservative (no mass to be accreted by the companion).
We checked the models with $M_{\rm 10}=1.20M_\odot$,
and found that the assumption of completely 
conservative mass transfer may significantly increase the parameter space for producing EL CVn stars (see Fig.19)
while that of completely non-conservation dramatically reduces the parameter space since the companion 
cannot increase in mass to become an A-or F-type dwarf star under this assumption. In our study, we
 adopted general assumptions as has been done
in previous studies \citep{chen02,chen09} and the produced EL
CVn-type stars have mass ratios consistent with observations
except for those with $P<1$ d, indicating that the assumption on
the amount of mass loss is in general reasonable. Some
more-evolved donors at the onset of mass transfer may have less
conservative evolution, and this may explain the observed
relatively high mass ratios for $P<1$ d. 

\begin{figure}
\includegraphics[width=6.0cm,angle=270]{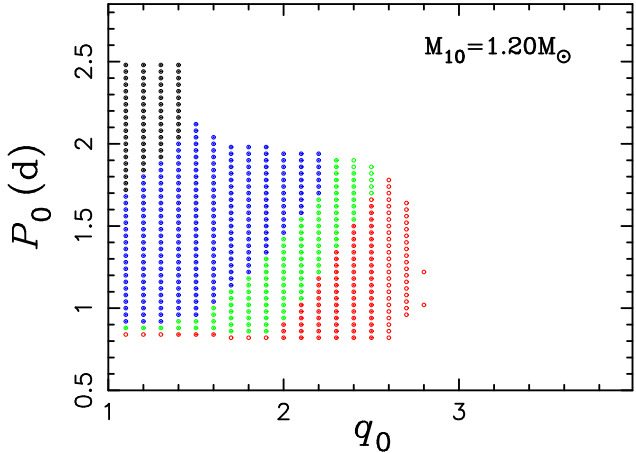}
\caption{Similar to Figs.3--5 but for the assumption of completely conservative mass tranfer.  
\label{120-con}}
\end{figure}

The influence of variable angular momentum loss can be estimated in the following way. In
each ($q_{\rm 0}, P_{\rm 0}$) plane, the upper boundary is
determined by the maximum mass transfer rate while the lower
boundary is related to $P_{\rm b}$. Although $P_{\rm b}$ changes
with assumptions in binary evolution, the lowest mass (or core
mass) for a given star to evolve through the constant-$L$ phase
does not change with any assumptions on binary evolution since
whether a star contracts (mass transfer terminates) or not is only
determined by its intrinsic structure, independent of detailed
mass loss process. So, more angular momentum loss in comparison to
that adopted in the calculations means a rapid shrinkage of the
Roche lobe and a faster mass transfer process. The star must then
be more evolved at the onset of mass transfer (a longer $P_{\rm
b}$) to evolve beyond the bifurcation period. As a consequence,
the lower boundary moves upwards. Meanwhile, the rapid shrinkage
of Roche lobe makes dynamically unstable mass transfer more likely
so the upper boundary moves downward. This effect might be small
since most potential progenitors of EL CVn-type stars obtained in
this paper have orbital periods just above the bifurcation period
and start mass transfer on the MS or in Hertzsprung Gap (HG),
where the mass transfer is generally stable unless we artificially
adopt an unreasonable large angular momentum loss. In general, a
large angular momentum loss leads to a smaller parameter space for
producing EL CVn-type stars, resulting in a lower birthrate and
fewer such objects. 

\subsection{Effects of Metallicity}
In this study we have only considered Pop I stars. Here we discuss
the expected influence of metallicity on the formation of EL CVn
binaries. Since the He cores in low-mass stars are very similar,
we may get several general impressions from the structure of the
envelope. (1) {\it The envelope mass $M_{\rm env}$}. For a given
core mass $M_{\rm c}$, a high $Z$ means a larger opacity in the
envelope, then a lower $M_{\rm env}$ when the star contracts to be
sure that the radiation pressure $P_{\rm r}$ is less than the
gravity pressure $P_{\rm grav}$ at that time. As a consequence,
the minimum mass (and the mass peak) increases with the decreasing
of $Z$. (2) {\it The lifetime of EL CVn-type stars, $t$}. As
discussion in section 3.2, $t \propto M_{\rm env}/\dot{M}_{\rm
nuc}$, where $\dot{M}_{\rm nuc}$ is very sensitive to the
temperature in the burning shell $T_{\rm b}$. Since $T_{\rm b}\sim
M_{\rm c}/R_{\rm c}$ \citep{ref70} and $R_{\rm c}$ decreases with
$Z$ for a given $M_{\rm c}$, the temperature $T_{\rm b}$, then
$\dot{M}_{\rm nuc}$, increase with decreasing $Z$. So, the value
of $t$ may not significantly deviate from that shown in this paper
since $M_{\rm env}$ also increases with the decreasing $Z$ as
discussed above. (3) {\it The parameter space for EL CVn stars.}
For a given ($M_{\rm 10}, q_{\rm 0},P_{\rm 0}$), a low $Z$
indicates a lower opacity, then a smaller stellar radius, which
means that, in comparison to a binary with the same parameters but
a high $Z$, the donor fills its Roche lobe at a later evolutionary
stage (or more evolved, since $R_{\rm L}$ is determined and not
affected by $Z$) and so is more likely to evolve through the
constant-$L$ phase. So, the bifurcation period, $P_{\rm b}$,
becomes smaller with decreasing $Z$, and the whole parameter
spaces move downwards.

\section{Conclusions}
In this paper, we systematically studied the properties of EL CVn
stars from the stable mass transfer channel, such as the
evolutionary tracks, the envelope masses, the lifetimes in the
nearly constant-$L$ phase, the component masses and the $M_{\rm
WD}-P_{\rm orb}$ relation. The results are summarized as follows.

(1) In general, the evolutionary tracks, mass ratios and the
$M_{\rm WD}-P$ relation are in good agreement with observations.
There is a small bulge at the low mass end on the
 $M_{\rm WD}-P$ relation due to non-degenerate cores of such low
mass donors at the termination of mass transfer. It starts from
$\sim 0.22M_\odot$ for $Z=0.02$ and $\sim 0.25M_\odot$ for
$Z=0.004$. The bulges may reduce the differences of the $M_{\rm
WD}-P$ relation induced by metallicity and makes the products from
various metallicities indistinguishable. Based on the $M_{\rm
WD}-P$ relation, the derived masses for the first two Kepler hot
WDs are consistent with that in previous studies, and all the
discovered EL CVn-type stars in \citet{maxted14a} have low-mass
($\le\sim0.2M_\odot$) proto-He WDs.

(2) The lifetimes in the nearly constant-$L$ phase $t$ are
strongly dependent on the masses of proto-He WDs $M_{\rm WD}$, i.e.
$t \propto M_{\rm WD}^{-8}$ for $M_{\rm WD}\le0.21M_\odot$ and
$t\propto M_{\rm WD}^{-10.5}$ for the lower lifetime limit. The
study gives $t\sim 2\times 10^7$ yrs for $M_{\rm WD}=0.25M_\odot$,
$\sim 4\times 10^8$ yrs for $M_{\rm WD}=0.2M_\odot$ and up to
$\sim 4\times 10^9$ yrs when $M_{\rm WD}=0.15M_\odot$. This leads
to an intrinsic peak mass around the minimum mass of proto-He WDs
($\sim 0.16M_\odot$), which is only determined by initial
composition and does not depend on details of the mass transfer
processes. The most common component masses of EL CVn-type stars
are $M_{\rm WD}=0.17-0.21M_\odot$ and $M_{\rm WD}=1.3-1.5M_\odot$
if selection effects are included, consistent with the
observations.

(3) The most likely progenitor mass for producing the observed EL CVn-type
stars (with orbital period less than 2.2 d) is in the range $1.15-1.20M_\odot$. 
Assuming an appropriate star formation rate for the Galaxy ($5M_\odot {\rm yr}^{-1}$) we
find that, the birthrate of EL CVn-type stars with shorter $P$ is
very small and the local density is less than $2\times 10^{-9}{\rm
pc}^{-3}$ for $P\le 2.2$ d in populations younger than 2.5 Gyr.
However, the role of EL CVn-type stars with short $P$ becomes more
and more important with age due to the very long lifetime, and the
number is $2-5\times 10^6$ for those with $P\le 2.2$ d in the
Galaxy, with a local density of $4-10\times 10^{-6}{\rm pc}^{-3}$. 
Though the number  or local density may change significantly with the assumptions
in modeling binary evolution, many more low-mass EL CVn-type stars are still expected and
they are in old populations more likely, as can be confirmed
observationally in future.

\section*{Acknowledgements}
We thank the anonymous referee for his/her very helpful suggestions on the manuscript.
This work is supported by the Natural
Science Foundation of China (Nos. 11422324, 11521303,
11390374), by Yunnan province (Nos. 2012HB037, 2013HA005) and by
the Chinese Academy of Sciences (No. KJZD-EW-M06-01).


\bibliographystyle{mnras}




\appendix
\section {The Common Envelope Ejection Channel}

Here we make a simple analysis on the common envelope evolution
(CEE) channel for the formation of EL CVn-type binaries. A
reasonable and commonly used approximation for CEE is based on
energy conservation. As a companion star spirals in a giant's
envelope, due to friction between the inner binary (the giant's
core and the companion) and the envelope, the orbital energy is
released and deposited in the envelope, and a part of the energy
may be used to eject the envelope. If we denote the binding energy
of the envelope as $\Delta E_{\rm B}$, the orbital energy released
during the spiralling process as $\Delta E_{\rm orb}$ and the
efficiency of energy conversion as $\alpha$, the envelope can only
be ejected when $\alpha\Delta E_{\rm orb}\ge\Delta E_{\rm B}$.
Furthermore, denoting the core and envelope masses of the giant
and its radius as $M_{\rm c}$, $M_{\rm e}$ and $R_{\rm 1i}$,
respectively, we have $\Delta E_{\rm B}=\frac{G(M_{\rm c}+M_{\rm
e})M_{\rm e}}{\lambda R_{\rm 1i}}$, where $G$ is the gravitation
constant and $\lambda$ is a dimensionless factor related to the
actual giant structure. In general, the companion star can hardly
accrete material in the CE process and we assume no mass to be
accreted by the companion. With this assumption, we have
\begin{equation}
\alpha [\frac{GM_{\rm c}M_{\rm 2i}}{2a_{\rm f}}-\frac{G(M_{\rm
c}+M_{\rm e})M_{\rm 2i}}{2a_{\rm i}}] \ge \frac{G(M_{\rm c}+M_{\rm
e})M_{\rm e}}{\lambda R_{\rm 1i}},
\end{equation}
where $M_{\rm 2i}$ is the mass of the companion and $a_{\rm i}$
and $a_{\rm f}$ are the orbital separation before and after the
CEE, respectively. The internal energy including ionization energy
\citep{han94} is also to be considered to play a role in the
ejection of the envelope, but the contribution is very small when
the giant is near the base of giant branch and is therefore
ignored here.

After multiplying Eq (A1) by the factor $\frac{\lambda R_{\rm
1i}}{G(M_{\rm c}+M_{\rm e})M_{\rm e}}$ and transposing the right
term of the equation to the left, we get
\begin{equation}
\frac{\alpha \lambda}{2}\frac{R_{\rm 1i}}{a_{\rm i}}\frac{M_{\rm
2i}}{M_{\rm e}}[\frac{M_{\rm c}}{M_{\rm c}+M_{\rm e}}\frac{a_{\rm
i}}{a_{\rm f}}-1]-1\ge 0.
\end {equation}

Defining the initial mass ratio as $q_{\rm i}=(M_{\rm c}+M_{\rm
e})/M_{\rm 2i}$ and the final mass ratio as $q_{\rm f}= M_{\rm
c}/M_{\rm2i}$, we have
\begin{equation}
\frac{M_{\rm 2i}}{M_{\rm e}}=\frac{1}{q_{\rm i}-q_{\rm f}},
\frac{M_{\rm c}}{M_{\rm c}+M_{\rm e}}=\frac{q_{\rm f}}{q_{\rm i}},
\end{equation}
and according to Kepler's third law, there is
\begin{equation}
\frac{a_{\rm i}}{a_{\rm f}}=(\frac{P_{\rm i}}{P_{\rm
f}})^{2/3}(\frac{1+q_{\rm i}}{1+q_{\rm f}})^{1/3},
\end{equation}
where $P_{\rm i}$ and $P_{\rm f}$ are orbital periods before and
after the CEE, respectively. The criterion for the ejection of the
CE is immediately written as
\begin{equation}
\frac{\alpha \lambda}{2(q_{\rm i}-q_{\rm f})}[\frac{q_{\rm
f}}{q_{\rm i}}(\frac{1+q_{\rm i}}{1+q_{\rm f}})^{1/3}(\frac{P_{\rm
i}}{P_{\rm f}})^{2/3}-1](\frac{R_{\rm 1i}}{a_{\rm i}})-1\ge 0.
\end{equation}
For convenience we denote the left part of Eq (A5) as $y$. The
criterion for CE ejection is then written as $y\ge0$. Note that
there is an additional constraint on the initial mass ratio, i.e.
$q_{\rm i}>1$, since the donor is the initially more massive
component from binary evolution.

At the onset of CEE, the giant just fills its Roche lobe, i.e.
$R_{\rm 1i}\simeq R_{\rm L}$, then we have
\begin{equation}
 R_{\rm 1i}/a_{\rm i}\simeq R_{\rm L}/a_{\rm i}=0.462(\frac{q_{\rm i}}{1+q_{\rm i}})^{1/3},
\end{equation}
where $R_{\rm L}$ is the Roche lobe radius of the giant (see
Paczy$\acute{\rm n}$ski 1971). Combining Eq (A6) with Kepler's
third law and $R_{\rm L}\simeq R_{\rm 1i}$ , we have the initial
orbital period \citep{rap95}
\begin{equation}
P_{\rm i}=20G^{-1/2}R_{\rm 1i}^{3/2}M_{\rm 1i}^{-1/2}
\end{equation}
where $M_{\rm 1i}=M_{\rm c}+M_{\rm e}$ is the donor's mass. Due to
the well-known $R_{\rm 1i}-M_{\rm c}$ relation for giants, i.e.
for Pop I stars \citep{rap95}
\begin{equation}
R_{\rm 1i}=5500M_{\rm c}^{4.5}(1+4M_{\rm c}^4)+0.5R_\odot,
\end{equation}
the value of $P_{\rm i}$ can be approximately given by $M_{\rm
1i}$ and $M_{\rm c}$. So we can obtain the value of $y$ from Eqs
(A6-8) and the observed $q_{\rm f}$ and $P_{\rm f}$, for given
values of $M_{\rm c}$, $\alpha \lambda$ and $q_{\rm i}$. It is
considered to be reasonable to suppose the value of $\alpha
\lambda$ to be in the range of $0.5-1$ \citep{rap15}. By setting
$\alpha \lambda=1.0$, which is an extreme case and gives the
maximum value of $y$, we examined a series values of $M_{\rm c}$
and $q_{\rm i}$ for all the 17 EL CVn-type stars in Table 2 of
\citet{maxted14a}. For each EL CVn-type sample, the constraints of
$y\ge0$ and $q_{\rm i}>1$ require $M_{\rm c}>0.3M_\odot$ (see Fig.
A1 for an example), far beyond the mass of proto-He WDs in EL
CVn-type binaries. This means that the CEE can not produce the
observed EL CVn-type stars.

\begin{figure}
\includegraphics[width=6.3cm,angle=270]{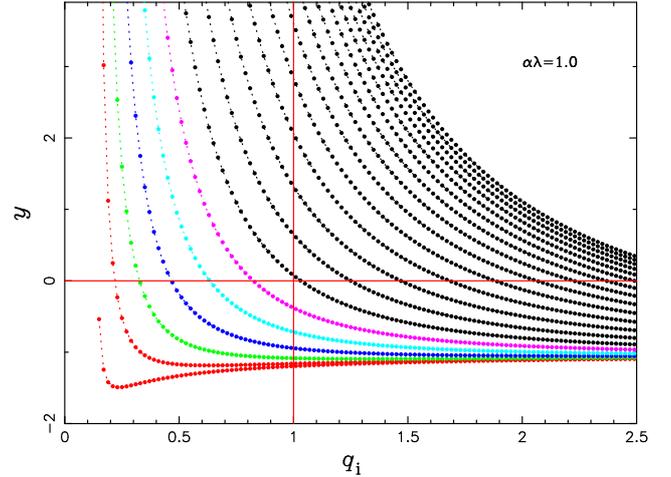}
\caption{The $(y, q_{\rm i})$ plane for WASP 2328-39, with orbital
period of 1.29 d and mass ratio of 0.13. The parameter $y$
indicates the left part of Eq (A5) and $q_{\rm i}$ is initial mass
ratio. The core mass $M_{\rm c}$ changes from 0.15 to $1.1M_\odot$
in steps of $0.05M_\odot$ from left to right. The common envelope
can only be ejected when $y\ge0$ (the horizonal line) and the
initial mass ratio should be greater than 1 (the vertical line)
from binary evolution. So, only the upper right part in the figure
is appropriate for the object produced from the CEE. To satisfy
both constraints we have $M_{\rm c}\ge\sim0.45M_\odot$ for this
object when $\alpha \lambda=1.0$, which is an extreme case and
gives the maximum value of $y$. \label{yplane}}
\end{figure}


\bsp	
\label{lastpage}
\end{document}